\documentclass[preprint]{aastex}
\usepackage{lscape} % Wes Custom
\usepackage{amsmath, amsthm,amssymb}
\usepackage{epstopdf}

\title{The Collisional Divot in the Kuiper belt Size Distribution}

\author{Wesley C. Fraser {1}}
\altaffiltext{1}{Division of Geological and Planetary Sciences, California Institute of Technology, MC 150-21, 1200 E. California Blvd. Pasadena, CA. 91125}

\date{} % delete this line to display the current date

\begin{abstract}
This paper presents the results of collisional evolution calculations for the Kuiper belt starting from an initial size distribution similar to that produced by accretion simulations of that region - a steep power-law large object size distribution that breaks to a shallower slope at $r\sim1-2$ km, with collisional equilibrium achieved for objects $r\lesssim 0.5$ km. We find that the break from the steep large object power-law causes a divot, or depletion of objects at $r\sim10-20$ km, which in-turn greatly reduces the disruption rate of objects with $r\gtrsim 25-50$ km, preserving the steep power-law behavior for objects at this size. Our calculations demonstrate that the roll-over observed in the Kuiper belt size distribution is naturally explained as an edge of a divot in the size distribution; the radius at which the size distribution transitions away from the power-law, and the shape of the divot from our simulations are consistent with the size of the observed roll-over, and size distribution for smaller bodies. Both the kink radius and the radius of the divot center depend on the strength scaling law in the gravity regime for Kuiper belt objects. These simulations suggest that the sky density of $r\sim1$ km objects is $\sim10^6-10^7$ objects per square degree. A detection of the divot in the size distribution would provide a measure of the strength of large Kuiper belt objects, and constrain the shape of the size distribution at the end of accretion in the Kuiper belt.
\end{abstract}

\begin{document}

\maketitle

\section{Introduction}

The Kuiper belt is a population of planetesimals outside the orbit of Neptune which exhibits high inclinations and eccentricities by many belt members \citep{Trujillo2001b,Brown2001}, and has a very low mass, $M \lesssim 0.1\mbox{ M$_\oplus$}$ \citep{Gladman2001,Bernstein2004,Fuentes2008}. The high relative encounter velocities, $v_{rel} \sim 1 \mbox{ km s$^{-1}$}$ \citep{Delloro2001} and infrequent collisions of the largest members \citep{Davis1997,Durda2000} make the growth of Eris-sized bodies impossible over the age of the Solar system. Accretion in the early stages of planet-building must have been in a more dense and quiescent environment allowing large objects to grow \citep{Stern1997a}. 

The size distribution of the Kuiper belt is the result of the accretionary and collisional processes that have gone on in that region, and therefore provides one of the main constraints on those processes. For a general review of the Kuiper belt size distribution, see \citet{Petit2008} \footnote{Much of the observations that constrain the size distribution for radii $r\sim20-60$ km are published in more recent works \citep{Fraser2009,Fuentes2009}.} There are three properties which describe the general shape of the Kuiper belt size distribution:
\begin{itemize}
\item the existence of the largest objects, Eris and Pluto
\item the large object size distribution for $D\gtrsim100$ km which is well characterized by a steep power-law $\frac{dN}{dR}\propto r^{-q}$ with $q\sim4.8$ \citep{Gladman2001,Petit2006,Fraser2008}
\item the existence of a ``roll-over'' at $r\sim25-60$ km where the size distribution flattens to a much shallower distribution than for larger objects \citep{Bernstein2004,Fuentes2008,Fraser2009}
\end{itemize}
\noindent
These three properties need to be reproduced by any successful attempt at recreating the accretionary and collisional history of the Kuiper belt region. While this has not yet been achieved, these properties provide some insight into the general history of the belt.

Accretion simulations such as those of \citet{Stern1996a,Kenyon2001} and \citet{Kenyon2002} have demonstrated that objects as large as Eris could have accreted on timescales shorter than the age of the Solar system if the mass of the Kuiper belt was at least two orders of magnitude larger than the current mass, and if the relative encounter velocities, and hence, the eccentricities and inclinations of the objects were significantly lower in the early Solar system (see review by \citet{Kenyon2008}). The existence of the steep large object size distribution however, suggests that accretion was a short-lived process \citep{Kenyon2002,Fraser2008}. Some event must have disrupted accretion - likely the same event which scattered the primordial Kuiper belt onto orbits with high inclinations and eccentricities as observed today (see a review of proposed processes by \citet{Morbidelli2008}) - otherwise the large object size distribution would be too shallow to be compatible with the observed distribution.

The simulations of \citet{Stern1996a,Kenyon2001} and \citet{Kenyon2002} cannot reproduce the large roll-over size. In these simulations, interaction velocities remain low, such that objects larger than $r\sim 1$ km do not experience disruptive collisions over the age of the Solar system. Thus, the size distribution for objects larger than $r\sim 2$ km exhibits a steep slope comparable to that observed today, and the accretion break, or roll-over, occurs at radii too small to be compatible with observations. This suggests that after the belt was dynamically excited, it remained massive enough for a time long enough to allow collisional erosion to produce the large roll-over observed today.

A model presented by \citet{Kenyon2004} could reproduce the large roll-over if gravitational stirring by Neptune at its current location was included early in the simulations. Accretion occurred at a sufficient rate in these simulations to produce Eris-sized objects. In these simulations, accretion lasted too long however, resulting in a large object size distribution too shallow to be compatible with observations. In addition, the mass loss due to collisional grinding was insufficient to produce the tenuous modern-day belt (see discussions by \citet{Morbidelli2008} and \citet{Kenyon2008}). The Kuiper belt must have undergone a more rapid excitation and mass depletion than occurred in these calculations.

\citet{Pan2005} presented an analytical, order of magnitude, collisional evolution model, reminiscent of the ground breaking work of \citet{Dohnanyi1969}. With this model, they demonstrated that the large roll-over size could be produced by collisional grinding on timescales shorter than the age of the Solar system. They however, assumed that the size distribution rolled over to collisional equilibrium. It has been demonstrated that there is a range of objects with sizes smaller than the roll-over,  which are preferentially eroded but which are not being replenished from fragments of larger disrupted objects \citep{Kenyon2002}. These objects do not achieve collisional equilibrium. Equilibrium is only achieved at a some smaller radius, the exact size of which, depends on the density of planetesimals. Thus, the model of \citet{Pan2005} likely, does not predict the correct shape of the size distribution smaller than the roll-over, or the rate at which the roll-over evolves to larger sizes.

\citet{Benavidez2009} present a collisional evolution model of the Kuiper belt, in which they include collisions from multiple dynamical classes. Using this model, they calculate the collisional evolution of the Kuiper belt, starting from an initially steep size distribution for all sizes, with slope similar to that observed for large objects. Their results confirm the findings of \citet{Pan2005}; collisional grinding with relative velocities comparable to that observed can disrupt the majority of objects with $r\sim50-100$ km, and in effect, produce a roll-over at that size. These calculations however, likely do not reproduce the collisional history of the Kuiper belt, as they start from an initial condition not expected from standard accretion processes \citep{Kenyon2002}. Accretion simulations which produce Eris-sized objects produce size distributions with an accretion break at $r\sim 1-2$ km, not a steep distribution for all sizes.

Here we present a collisional evolution model, which we utilize to calculate the size distribution of the Kuiper belt from some starting condition. With this model, we wish to determine whether or not collisional erosion could produce the $r\sim20-50$ km roll-over, starting from a size distribution similar to that produced by models of accretion in the outer Solar system, but in the dynamically excited conditions of the current day Kuiper belt. From these calculations, we wish to constrain the shape of the modern day size distribution smaller than the roll-over, and to constrain the mass of the Kuiper belt at the end of accretion.

In Section~\ref{sec:model} we present our collisional evolution model, and the planetesimal strength and shattering models we adopt. In Section~\ref{sec:results} we present the main results of our calculation, namely that that existence of a break at $r\sim1$ km left-over from accretion causes a divot in the size distribution at larger sizes. In Section~\ref{sec:discussion} we present the consequences of our model, and finish with concluding remarks in Section~\ref{sec:conclusions}.

\section{The Model \label{sec:model}}
Our collisional evolution model uses a formalism similar to that of \citet{Ohtsuki1990} which uses bins of constant mass, but tracks average bin radius\footnote{Bins defined by an average radius rather than mass were adopted to ease analysis of the resultant size distributions. This choice however, has no effect on the results.}. The model considers a swarm of planetesimals distributed in a finite number of size bins, where bin $i$ contains a number of objects, $N_i$, with radii $r_i<r<r_i+\delta r_i$, where $\delta\sim1.1$.\footnote{Note that $\delta$ here is the cube-root of the usual "mass-delta" - the ratio of masses in consecutive bins - in other calculations \citep[see][]{Kenyon1998}} In our model we assume that all particles have the same relative collision velocity, $v_{rel}$. We do not include velocity evolution in our model, and assume that $v_{rel}$ is a constant during the simulations. Each collision can either result in catastrophic disruption and dispersal if the collision energy, $Q$, is sufficiently high, or cratering and mass accretion otherwise. We adopt the standard center-of-mass collision energy $Q=\frac{1}{4} \frac{m_i m_j v_{rel}}{m_i+m_j}$. In the following sections we describe the details of the model and our prescription for the collisional outcomes.

\subsection{Disruption and Cratering Model \label{sec:disruptionModel}}
We adopt the catastrophic disruption energy threshold functional form presented by \citet{Benz1999}. This form includes contributions from the tensile strength of the material body, as well as a gravitational binding term, and is given by

\begin{equation}
Q_D^*=Q_o\left(\frac{r}{1 \mbox{ cm}}\right)^a+B\rho\left(\frac{r}{1 \mbox{ cm}}\right)^b
\label{eq:Qdstar}
\end{equation}

\noindent
where $Q_D^*$ is defined as the disruption energy per unit mass of the target body required to shatter and disperse 50\% of the target into a spectrum of fragments. $\rho$ is the target density which we assume is constant for all sizes, and $Q_o$, $a$, $b$, and $B$ are constants appropriate for the material properties of Kuiper belt objects. 

\citet{Benz1999} utilized smooth particle hydrodynamic simulations to determine the outcome of disruptive collisions between icy bodies. They calibrated Equation~\ref{eq:Qdstar} under a large range of impact parameters (velocity, size ratio, etc.) for both basalt and water ice targets. They found that the disruption energy and hence the constants $Q_o$, $a$, $b$, and $B$  depended on the relative impact velocity. They also found that the mass of the largest remaining fragment $M_{LRF}$ depended linearly on the ratio of the impact energy to the disruption energy, $Q/Q_D^*$. They demonstrated that the mass of the largest remaining fragment can be well represented by

\begin{equation}
\frac{M_{LRF}}{M_{t}} = \gamma = Y-X\left(\frac{Q}{M_{t}Q_D^*}\right) 
\label{eq:MLRFMT}
\end{equation}

\noindent
where $M_{t}$ is the initial target mass. Examination of Figure 9. from \citet{Benz1999} reveals that $Y=1.0$ and $X=0.55$ in Equation~\ref{eq:MLRFMT} is an acceptable representation of $\gamma$, and we adopt these parameters for our calculations. \citet{Stewart2009} have confirmed the findings of \citet{Benz1999} and from their simulations have determined that $X\approx0.48$. The small difference in $X$ found by \citet{Stewart2009} and the value we adopt here will make an insignificant difference on our results, and will have no effect  on our conclusions.

%In our simulations we adopt the strength parameters presented by \citet{Benz1999} for collisions of water-ice bodies with relative velocities of $v_{rel}=0.5$ km/s, which is similar to the true impact velocities of Kuiper belt objects \citep{Delloro2001}. \citet{Leinhardt2009} however, have demonstrated that the destruction criterion of \citet{Benz1999} is likely a factor of $\sim3$ too high for objects in the Kuiper belt. In our calculations we vary the strength parameters of Equation~\ref{eq:Qdstar} to determine the effect of the uncertain disruption criterion on our simulation results.

We assume that the distribution of collisional fragments from a disruption is a power-law, and is given by $dN/dr = Ar^{-q_D}$ where $q_D$ is the logarithmic slope. It has been shown that in the asteroid belt, the distribution of fragments is better represented by a two slope distribution \citep{Bottke2005}. In our calculations we considered a range of two sloped models. We found that, for a reasonable range of two-sloped models, the change in the size distribution results were small compared to the changes caused by adjustment of the strength and cratering laws - also much less than the variation introduced into the results from the unknown collisional history. Thus, we choose the power-law as a simple, scale free, one parameter representation that avoids including much of the uncertain physics of fragmentation in our model.

Assuming that the density of planetesimals is constant for all sizes, from Equation~\ref{eq:MLRFMT} we see that the radius of the largest remaining fragment is given by

\begin{equation}
r_{LRF} = r_t \gamma^{1/3}
\label{eq:rlrf}
\end{equation}
\noindent
where $r_t$ is the radius of the target body. Assuming there is only one largest remaining fragment, the normalization of the fragment model is given by $A=(q_D-1) \gamma^{(q_D-1)/3} r_t^{(q_D-1)}$ where $q_D>1$. Conserving mass, we have

\begin{equation}
\int_0^{r_{LRF}} A r^{-q_D} m(r) dr = m(r_t)
\label{eq:mass_cons}
\end{equation}

\noindent
where $m(r)$ is the mass of an object with radius $r$. Equation~\ref{eq:mass_cons} can be solved for $q_D$ to give 

\begin{equation}
q_D=\frac{\gamma+4}{\gamma +1}.
\end{equation}
\noindent
As $\gamma>0$, for our fragment model, $1<q_D<4$.

For impacts with energy insufficient to disrupt the target, we considering a cratering model where the excavated crater mass is given by

\begin{equation}
M_{Crat}=\alpha f_{frac} Q
\label{eq:Mcrat}
\end{equation}

\noindent
where $\alpha$ is the crater excavation coefficient, which depends on the material properties of the target and impactor, and $f_{frac}$ is the fraction of kinetic energy that went into shattering the target material. We consider values of $\alpha=10^{-8}-10^{-9} \mbox{ s$^2$ cm$^{-2}$}$ \citep{Petit1993}. 

We assume that the mass fraction of ejected material with velocity greater than $v$ is given by 
\begin{equation}
f_M(v)=\left(\frac{v}{v_{min}}\right)^{-k}
\label{eq:Vfrac}
\end{equation}
\noindent
where $k=\frac{9}{4}$ \citep{Petit1993}. By assuming that a fraction, $f_{KE}$ of the impact energy is imparted into kinetic energy of the excavated material (Equation~\ref{eq:Mcrat}), and requiring conservation of energy, the minimum ejecta velocity is given by $v_{min}=\sqrt{\frac{2}{9} \frac{f_{KE}}{\alpha f_{frac}}}$, with the fragment mass escaping the body given by $M_{esc}=M_{Crat}\left(\frac{v_{esc i,j}}{v_{min}}\right)^{-k}$, where $v_{esc\mbox{} i,j}=\sqrt{\frac{2G(m_i+m_j)}{r_i+r_j}}$ is the mutual escape velocity of colliding pair $i$, $j$ \citep{Wetherill1993}. Clearly, accretion will occur if $M_{esc}$ is smaller than the mass of the impactor.

We assume that the crater fragment distribution is a power-law with slope $q_c=-3.4$, and a largest remaining fragment 

\begin{equation}
r_{LRF,c}=\left(\frac{4+q_c}{4\pi\rho} M_{Crat}\right)^{1/3}.
\label{eq:CLRF}
\end{equation}

Simulations of \citet{Benz1999,Leinhardt2009} and \citet{Stewart2009} have demonstrated that catastrophic collisions still occur for energies as little as $Q\sim 0.2 \mbox{ } Q_d^*$. If we consider a maximum crater size as a fraction of the target body radius, $f_{Crat}$, and require continuity between cratering and disruption regimes, the energy threshold for which collisions are {\it just} disruptive, as a fraction of $Q_D^*$, is given by $f_{dis}=\frac{1}{X}\left(1- (1-f_{Crat})^3\right)$. A minimum disruption energy criterion $f_{dis}=0.2$ corresponds to $f_{Crat}=0.04$. In this way, events which strip more than $\sim10\%$ of the mass of the target are considered disruptive collisions, while those that strip a smaller amount, are considered cratering events. In our calculations, we consider maximum crater sizes $f_{Crat}=0.05-0.15$ corresponding to minimum catastrophic disruption energy thresholds of $0.4-0.7 Q_D^*$.

We note here that there are many effects other than size that determine disruption strength and collisional outcome, such as impactor size and target porosity (see discussion by \citet{Leinhardt2008}). Given the uncertainty in the knowledge of the processes that shaped the Kuiper belt size distribution, the specifics of these effects are of little import, and we do not consider them in the collisional model given by equations~\ref{eq:Qdstar}-\ref{eq:CLRF}.

\subsection{Collision Rate}
From simple cross-section arguments, it can be shown that, if objects from bin $i$ are passing through the swarm of objects from bin $j$, the number of collisions between the objects from bins $i$ and $j$, in time-step $\Delta t$ is

\begin{equation}
N_{coll_{i,j}}=\frac{\pi ( r_i^2 +r_j^2) v_{rel} \Delta t}{V} N_iN_j
\label{eq:Ncoll}
\end{equation}

\noindent
where $v_{rel}$ is the relative encounter velocity, and $V$ is the volume occupied by the swarm of particles. 

Because a single disruptive collision, or accretion of that object onto a larger target will remove an object from a bin, Equation~\ref{eq:Ncoll} does not represent the true number of collisions between bins $i$ and $j$ - there can only be as many disruptive collisions as there are objects in a bin to be disrupted. Therefore, in calculating collision rates, one must consider disruptions of bin $i$ by other impactors $k\neq j$, or if $i$ is accreted onto a larger object. We adopt a probabilistic approach. We set the probability that an object from bin $i$ is disrupted by an object from bin $j$ (or accreted onto $j$ if $r_j>r_i$), $P_{i,j}$, as the probability that the disruption of the target $i$ was from an impactor from bin $j$, $P_{D_{i,j}}$, times the probability that an object from bin $j$ was available to cause the disruption, $P_j$.

$P_{D_{i,j}}$ is simply given by the number of collisions of objects from bin $i$ by those from bin $j$, given by Equation~\ref{eq:Ncoll}, divided by the number of objects in bin $i$. That is $P_{D_{i,j}} = \frac{N_{coll_{i,j}}}{N_i}$.

The probability that impactor $j$ is available can be found by setting $P_j = 1-\bar{P_j}$ where $\bar{P_j}$ is the probability that an object $j$ is disrupted by or accreted by an object with radius $r_k\neq r_i$. That is, $P_j = 1-\frac{1}{N_j}\sum_{k\neq i} N_{coll_{j,k}}$. The true number of disruptive collisions from bin $i$ by bin $j$ is then

\begin{equation}
N_{disrupt_{i,j}} = N_i P_{D_{i,j}} P_j.
\label{eq:Ndis}
\end{equation}

\noindent
Equation~\ref{eq:Ndis} predicts a smaller number of collisions for large objects compared to Equation~\ref{eq:Ncoll}. The difference imposses significant changes in the long-term collisional evolution when the number of disruptive collisions in a time-step is $\gtrsim1\%$ the number of objects in that bin. If the time-step is small enough, then the effect is small.

For completeness we apply a corrective term to Equation~\ref{eq:Ncoll}, $\left(1+\left(\frac{V_{esc i,j}}{V_{rel}}\right)^2\right)$ to account for gravitational focusing.  For the impact velocities considered in this work, this term is negibibly small, and could be ignored without changing our conclusions.

\subsection{Code Details and Nominal Parameters \label{sec:details}}
In any given time-step, the number of disruptive collisions per bin is first calculated with Equation~\ref{eq:Ndis}, using an adaptive time-step, such that the number of objects being removed from a bin - either from disruption, or accretion and cratering - is no more than 1\% the number of objects in the bin before the time-step. This minimizes code run-times while preserving the correct collisional evolution (see Appendix A). If the probability of collision from Equation~\ref{eq:Ndis} is less than unity, then a single collision occurrence is decided randomly. ie. the collision occurs if $R< N_i P_{Di,j} P_j$, where $R$ is a randomly generated number with $0\leq R\leq1$. 

Formally, if the number of collisions of a bin is less than $10^8$ per time-step, the number of collisions is kept as an integer value, and a random number generator is used to determine if the number is rounded up or down. \citet{Wetherill1989} has demonstrated that requiring an integer number of collisions per time-step when the number is less than $10^8$ is a sufficient condition to produce the correct time averaged collision rates.

For each non-disruptive collision between a large object $i$ and a small object $j$, the resultant body has a mass $m(r_i)+m(r_j)-M_{esc}$. An object is removed from bin $j$, and if the resulting mass is sufficiently large, a body is removed from bin $i$ and added to the bin with the appropriate bin $k$. For these collisions, their is a difference in mass between the resultant body, and the average mass of the bin it was promoted to. This mass, accreted by bin $k$, $M_{acc,k}$ (negative if cratering occurs) is accumulated over multiple time-steps. When the accreted (or cratered) mass becomes large enough, objects can ``grow'' (or shrink) to subsequent bins, ie. when $M_{acc,k}>M_{k\pm1}-M_k$ (sign set according to accretion or cratering) objects are moved randomly from bin $i$ to $k\pm1$. This is done in such a way as to reflect the fact that the accreted mass is occurring over all objects in bin $k$ and not just one. Thus, $\frac{M_{acc,k}}{|M_{k\pm1}-M_k|}$ objects are added to bin $k\pm1$ from bin $k$ when $\frac{M_{acc,k}}{N_k}>R$ where $R$ is a randomly generated value with $0\leq R \leq1$. This process preserves total system mass.

We consider bins increasing logarithmically in radius. That is $r_i=r_{i-1}\delta$ where $\delta>1$. For our standard simulations, we consider $\delta=1.1$. It has been shown that a finite $\delta$ produces an acceleration in the results (\citet[see][]{Wetherill1990}, and Appendix A). THus, our conclusions drawn here have the caveat that the evolution might occur in nature slightly slower than in our simulations.

\citet{CampoBagatin1994} has demonstrated that a finite minimum size bin will cause substantial wave-like deviations of the small size distribution away from the expected collisional equilibrium distribution; the smallest particles have no smaller impactors, and are in over abundance compared to larger particles. This over abundance creates an increased collision rate for larger particles, and so on. To prevent such a behavior, we enforce a collisional equilibrium slope for the smallest bins by removing excess objects from those bins. While this method is not the most accurate way of correcting for the finite bin minimum \citep[see][for a discussion]{Kenyon2008}, it is trivial to implement, and ensures accurate enough small object treatment to not affect our conclusions about the $\gtrsim 1$ km size distribution.  It was found that enforcing collisional equilibrium over the smallest $30-40$ bins is necessary to prevent the over abundance of the smallest objects (see Appendix \ref{sec:AppendixA}). 

Our calculations were performed on an NVIDIA GTX 280 graphics card using the CUDA programming environment \footnote{www.nvidia.com/cuda}. These cards provide 30 processors, each containing 8 processor cores, all running in parallel, and excel at highly parallel calculations, with speeds up to 50 times higher than typical multi-core CPUs. Due to the nature of the calculations however, we are limited to $2^n$ bins. For our nominal calculations, we use 256 bins with $\delta =1.1$ for our nominal simulations, allowing us to probe more than 10 orders of magnitude in radius. We considered larger $\delta$ with fewer bins to determine the effects of bin width and minimum bin size on our results. We found that changing the bin size did not produce a noticeable affect on our conclusions considering the range of results caused by the uncertain shattering and cratering models. Thus, we adopt $\delta=1.1$ for all the calculations we present here.

We wish to model collisional evolution that occurs in dynamical conditions typical of the modern day Kuiper belt. To that extent, we roughly model the Kuiper belt as a single annulus, with inner and outer radii of $30$ and $60$ AU. We set a scale height of $20$ AU similar to the scale height of the current Kuiper belt. We note that this scale height is significantly larger than that achieved in standard accretion calcuations \citep[see][for discussion]{Kenyon2002}. Such a scale height can only be achieved from scattering dynamics not included in those accretion models, and which is necessary to shape the primordial Kuiper belt to its current state \citep[see][]{Morbidelli2008}.

It has been shown that accretion calculations which utilize a single annulus, as done here, produce different results than those which utilize many annuli \citep{Kenyon2002}. The calculations we present here are intended to show the collisional evolution of an excited Kuiper belt. Standard multi-annulus codes cannot accurately represent the dynamical structure of the Kuiper belt and therefore do not accurately model the collisional evolution of this region \citep{Benavidez2009}. Rather, a proper treatment would involve a combined N-body and collisional evolution calculation, as done in \citet{Bottke2005} and \citet{Charnoz2007}. Given that the dynamical history of the Kuiper belt is not yet known, such a calculation is beyond the scope of this work. As such, the calculations presented should only be interpreted as rough approximations to the true collisional evolution that occurred in the Kuiper belt.

We set $v_{rel} = 1\mbox{ km s$^{-1}$}$  which \citet{Delloro2001} has demonstrated is a typical collision velocity in the modern Kuiper belt. At this velocity, targets of $r\gtrsim5$ km will be disrupted in collisions with size ratio larger than $\sim1:10$ for the impactor and target. Collisions with smaller impactors will typically cause cratering, and accretion will occur - slowly - for only the largest  ($r\gtrsim 250$ km) objects.

For our simulations we define two strength models. For the ``strong" model, we adopt the strength parameters of water-ice for a $0.5 \mbox{ km s$^{-1}$}$ collision velocity presented by \citet{Benz1999}, with $Q_o=7\times10^7 \mbox{ erg g$^{-1}$}$, $B=2.1 \mbox{ erg cm$^3$ g$^{-2}$}$, $a=-0.45$, and $b=1.19$. \citet{Leinhardt2009} however, has demonstrated that the above destruction criterion is likely a factor of $\sim3$ too high for objects in the Kuiper belt. Thus, we adopt the strength parameters suggested by \citet{Leinhardt2009}, which have the same slopes $a$ and $b$, as the strong model, but with a factor of 3 less disruption energy at all sizes for our ``weak'' model. We utilize the strong model for our nominal simulations, and discuss the effects of using the weak model in Sections~\ref{sec:results} and \ref{sec:discussion}.

We set the maximum crater size, $f_{Crat}=0.05$, and partition the collision energy into fracturing and kinetic energy as $f_{frac}=0.8$ and $f_{KE}=0.1$. Following \citet{Petit1993}, for our nominal model, we set the cratering velocity fragment $k=\frac{9}{4}$, the cratering efficiency $\alpha=10^{-9} \mbox{ s$^2$ cm$^{-2}$}$ corresponding to ``strong'' surfaces or low cratering efficiency, and the fragment size distribution slope $q_c=3.4$.

%For the nominal simulations, we consider the single power-law disruptive fragment size distribution presented in Section~\ref{sec:disruptionModel}. We performed a set of simulations with a broken power-law fragment distribution similar to that presented in \citet{Bottke2005}. We found that, for a reasonable range of parameters for the broken power-law fragment distribution, the size distribution results changed by at most 10\% from bin to bin compared to the same calculations done with the single sloped model we adopt here. The range of uncertainty on the strength and cratering models causes a variation in the results significantly larger than the use of the more realistic (and more complicated) fragment model. Therefore, the use of fragment models with a larger number of free parameters was not warranted for this work, and we adopt the single slope model here.

%To understand the dependence of the simulation results on our fragment model, we also consider a broken power-law fragment distribution like that presented in \citet{Bottke2005}, that breaks from slope $q_{D1}$ to $q_{D2}<q_{D1}$ at a radius $r=f_B r_{LRF}$, where $f_B\sim0.05$ and $q_{D1}\sim 4-6.5$. By conserving conservation of mass, and setting the number of largest remaining fragments to 1, the slope $q_{D2}$ can be solved providing a two-parameter fragment distribution. 

\subsection{Initial Conditions}
The goal of this work is to determine the collisional evolution of the Kuiper belt size distribution after the epoch of accretion. To that end, we consider a ``typical'' size distribution found from accretion simulations \citep{Kenyon1998,Kenyon2001,Kenyon2004,Stern1996a,Stern1997a}. That is, one which has a steep power-law for the largest objects with  $q_1\gtrsim4$. At some smaller size, $r_{b1}$, the slope becomes shallower, or breaks to $q_2\sim -1\mbox{ to } 1$ \citep{Stern1996a,Stern1997a,Kenyon2002,Kenyon2004} - we utilize a sharp break for simplicity. The ``break-radius'' $r_{b1}$ - typically 2 km for the Kuiper belt \citep{Kenyon2002}, corresponds to the size that objects which are smaller have undergone at least one disruptive collision. At some even smaller size, $r_{b2}\sim 0.5$ km, collisional equilibrium is reached, and the size-distribution slope turns up to the canonical $q_3=3.5$. Again we assume a sharp slope-change. The collisional equilibrium slope typically only deviates at sizes small enough such that radiation effects remove objects from the belt \citep{Thebault2003}. The effect however, is inconsequential on our results, so we do not consider radiation effects here.

The most recent measurements of the Kuiper belt size distribution have determined that $q_1\sim4.8$ \citep{Fuentes2008,Fraser2009} and breaks at a radius $r_{b1}\sim20-60\mbox{ km}$. The steep slope implies a short accretion timescale \citep{Kenyon2002,Fraser2008} and has been preserved from the epoch of accretion \citep{Durda2000,Pan2005}. Thus, for our nominal simulations, we start with an initial large object slope equal to that observed in the modern day belt, ie. $q_1=4.8$. We start with break radii and slopes inspired by the simulations of \citet{Kenyon2002} and \citet{Kenyon2004} which represent the most complete simulations of accretion in the Kuiper belt range to-date, and for our nominal simulations we take $q_2=0.0-2.0$, $q_3=3.5$, $r_{b1}=2$ km, and $r_{b2}=0.5$ km. We discuss variations of these initial conditions in Section~\ref{sec:discussion}.

Accretion simulations of the region have demonstrated that the Kuiper belt must have had a few orders of magnitude more mass \citep[see][for a review]{Kenyon2002} than currently observed \citep{Trujillo2001b,Gladman2001,Bernstein2004,Fuentes2008}, otherwise Pluto could not have achieved its size over the age of the Solar system. The exact initial mass available at the late stages of accretion however, is unknown. For our nominal simulations we adopt a $45 \mbox{ M$_{\oplus}$}$ Kuiper belt and vary this mass to test the evolution for different planetesimal densities. We note here that $45 \mbox{ M$_{\oplus}$}$ is at the large end of the mass thought to have existed in the primordial Kuiper belt \citep{Kenyon2002, Gomes2003}. This, the rate of evolution produced in calculation with such a high mass should be interpreted as upper limits to the real evolution. The parameters for our nominal simulations are listed in Table~\ref{tab:params}.

\section{Results \label{sec:results}}
\begin{figure}[h] %  figure placement: here, top, bottom, or page
   \centering
   \epsscale{0.85}
   \plotone{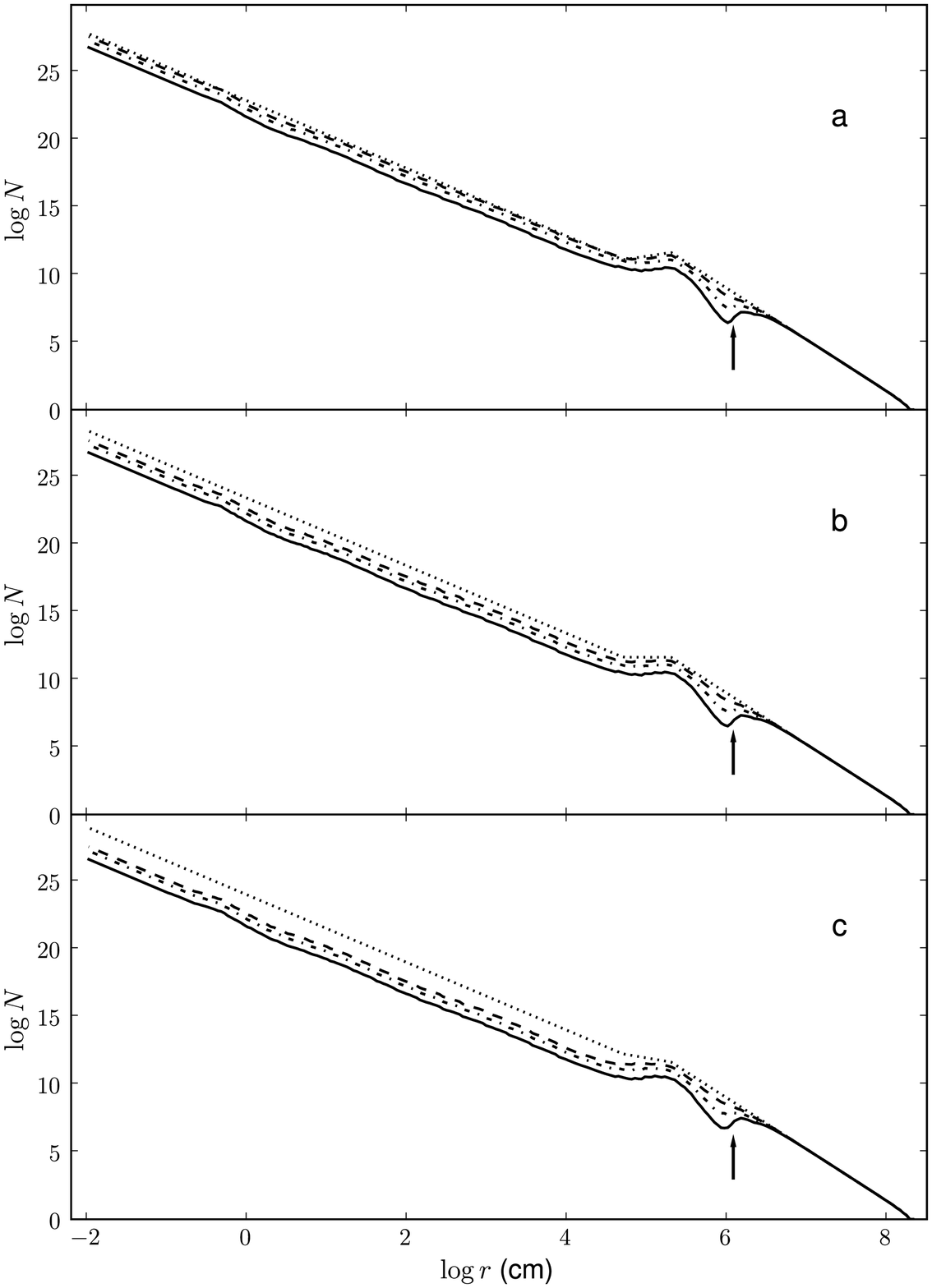}
   \figcaption{Results of the nominal simulations with the accretion break, $r_{b1}=2$ km at different times in the evolution. \textbf{Dotted line} - initial distribution. \textbf{Dashed line} - 25 Myr. \textbf{Dashed-dotted line} - 100 Myr. \textbf{Solid line} - 500 Myr. \textbf{a} $q_2=0$. \textbf{b} $q_2=1$. \textbf{c} $q_2=2$. The arrows mark the approximate divot center radius $r_{div}$ predicted from Equation~\ref{eq:divot}. NOTE: The radius bin-widths increase proportionally with the bin-centers. This correctly presents the number of objects in a bin, but causes the size distribution slopes to appear $-1$ shallower than the true slope. \label{fig:fig1}}   
\end{figure}

In Figure~\ref{fig:fig1}, we present the results of our nominal simulations at specific times. The expected behavior of the collisional erosion occurs at the large and small scales. For particles smaller than the initial $r_{b2}$, because these particles were initialized with the collisional equilibrium slope, the overall trend is to remain with that slope, and to decrease in mass as more particles are ground away.  For the largest ($r\gtrsim150$ km) objects, very few (if any) disruptive collisions occur. Accretion and cratering results in changes in the large object size distribution by less than $\pm \sim1 \%$ over the age of the Solar system.

For objects with $r\sim 10\mbox{ } r_{b1}$, a distinct divot forms in the size distribution. This behavior is caused by the initial break or turn-over at $r_{b1}$; the decrease in slope causes an absence in the number of disruptors available to shatter objects with $r\sim r_{b1}$. Thus, due to the change in slope at $r_{b1}$, any objects capable of being shattered by impactors with $r\sim r_{b1}$ experience an enhanced collisional disruption rate. This in turn reduces the disruption rate of objects capable of being shattered by the impactors with $r\sim10\mbox{ }r_{b1}$, which produces a distinct kink where the size distribution rolls away from the initial accretion slope $q_1$ into the divot. 

The location of the divot can be roughly estimated by equating the collision energy of impactors at the accretion break radius $r_{b1}$ to the disruption energy of the bodies with the divot radius $r_{div}$. Ignoring the tensile strength term in Equation~\ref{eq:Qdstar}, and solving for $r_{div}$  we find

\begin{equation}
r_{div}=\left(4 f_{dis} B \rho\right)^{\frac{-1}{3+b}} \mbox{ }  v^{\frac{2}{3+b}} \mbox{ } r_{b1}^{\frac{3}{3+b}}.
\label{eq:divot}
\end{equation}
\noindent

%In our discussion of the results, we define the ``kink'' radius, $r_{k}$, to be the radius  at which the size distribution rolls away from the steep accretion slope into the divot. $r_k$ can be estimated by replacing $r_{b1}$ with the divot center, $r_{div}$, in Equation~\ref{eq:divot} (see Figure~\ref{fig:fig1}).

\begin{figure}[h] %  figure placement: here, top, bottom, or page
   \epsscale{}
   \centering
   \plotone{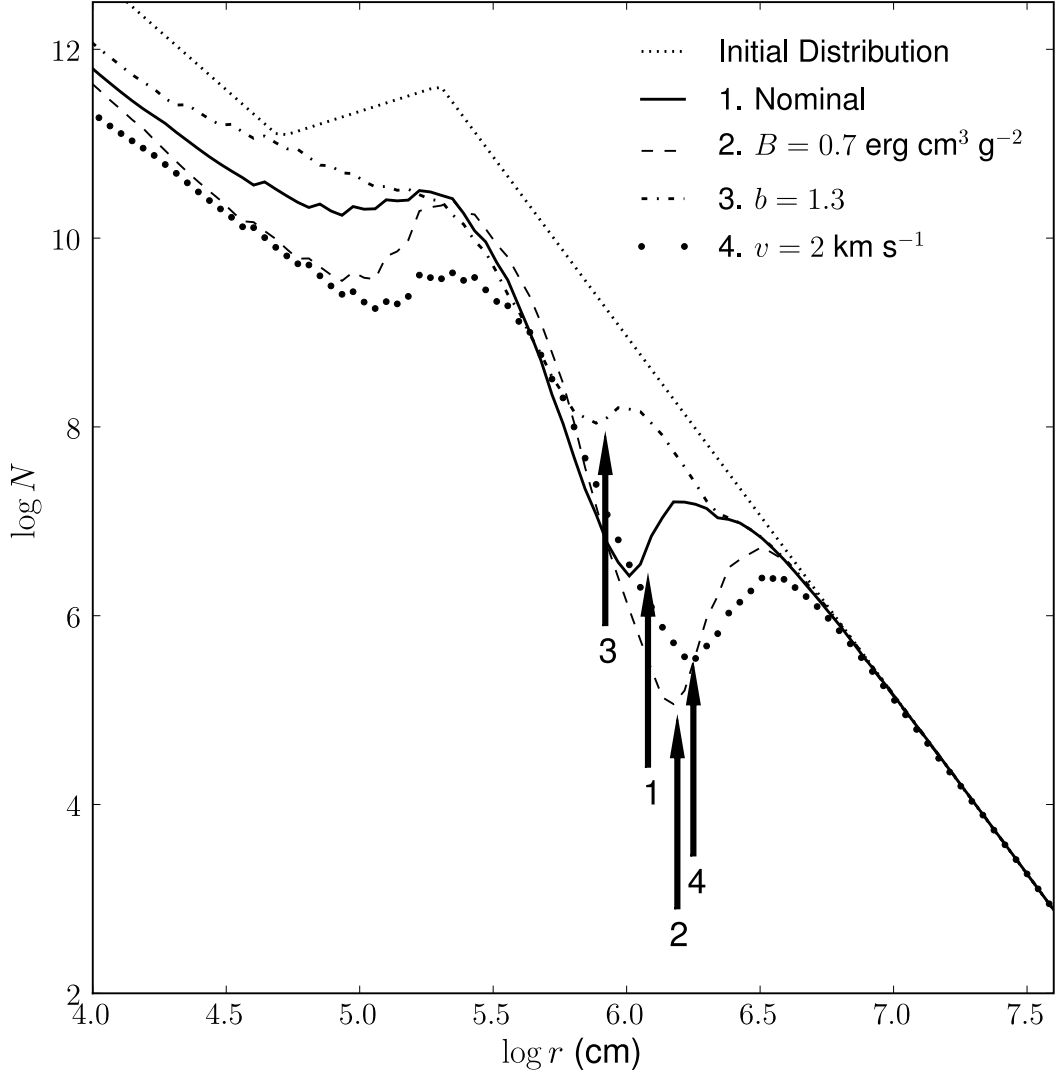} 
   \figcaption{Results of the simulations with the accretion break, $r_{b1}=2$ km at 500 Myr but varying parameters relevant to the location of the divot radius $r_{div}$. The arrows mark the approximate divot radii predicted from Equation~\ref{eq:divot}. \label{fig:fig2}}   
\end{figure}

Equation~\ref{eq:divot} demonstrates that given the accretion break size $r_{b1}$, the radius of the divot, $r_{div}$, depends on the strength parameters of the gravity term in Equation~\ref{eq:Qdstar}, $B$, and $b$, the density of Kuiper belt objects, $\rho$, the average interaction velocity, and the energy at which impacts transition from the cratering regime to the disruption regime, given by $f_{dis}$. The factor $\left(4 f_{dis} B \rho\right)^{\frac{-1}{3+b}}$  however, is of order unity. Thus, the central divot radius depends mainly on the slope of the gravity strength scaling-law $b$, and the velocity at which objects interact. These conclusions are confirmed in Figure~\ref{fig:fig2}, where we present the results of our nominal simulations with different strength parameters.

The rate at which the divot deepens depends on the slope change at $r_{b1}$; the greater the difference between $q_1$ and $q_2$, the greater the rate at which the divot deepens (see Figure~\ref{fig:fig1}). The divot formation timescale can be roughly estimated from Equation~\ref{eq:Ncoll}. The divot is formed primarily from catastrophic collisions of objects larger than the initial break radius $r_{b1}$ where the initial size distribution is given by a power-law with slope $q_1$, ie. $\frac{dN}{dr}=Cr^{-q_1}$ where $C$ is a normalization constant. Thus, by inserting a power-law for the number of smaller objects $N_j$ in Equation~\ref{eq:Ncoll}, integrating over all sizes of object capable of disrupting objects at the divot, ie. from $r_{b1}$ to $r_{div}$, and solving for time, we find that the timescale for divot formation $t_{div}$ is given by

\begin{equation}
t_{div} = \left(\frac{N_c}{N_{r_{div}}} \right) \frac{V}{\pi v C r_{div}^{3-q_1}}\left[ \frac{1-f^{1-q_1}}{1-q_1} + \frac{1-f^{3-q_1}}{3-q_1}\right]
\label{eq:tdiv}
\end{equation}

\noindent where $\frac{N_c}{N_{r_{div}}}$ is the fraction of objects with radius $r_{div}$ that have been disrupted, and $f=\frac{r_{b1}}{r_{div}}\sim 0.2$ (see Equation~\ref{eq:divot}). By adopting the nominal parameters, and saying a divot has formed when 99\% of objects at the divot have been disrupted, ie. $\frac{N_c}{N_{r_{div}}}=0.99$, then the divot formation timescale is $t_{div} \sim 25$ Myr. For this timescale estimate, we have ignored disruptions of objects with radii $r_{b1}\leq r\leq r_{div}$. Thus, Equation~\ref{eq:tdiv} represents a lower-limit of the true divot formation timescale. The true divot formation timescale is $\sim50\%$ longer if the break slope is $q_2=0$ and $\sim3$ longer if  $q_2=2$ (see Figure~\ref{fig:fig1}).

\begin{figure}[h] %  figure placement: here, top, bottom, or page
   \centering
   \plotone{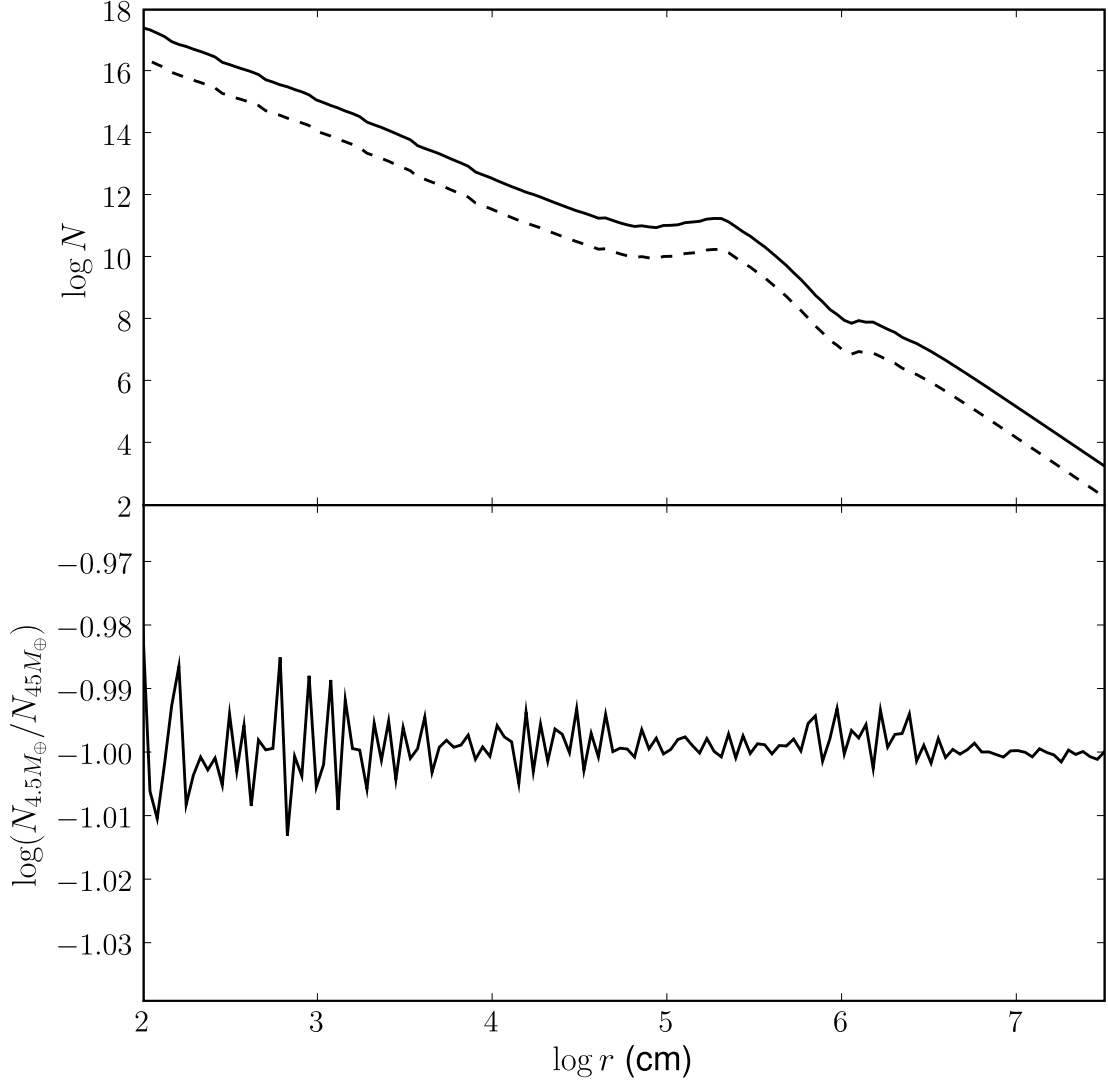} 
   \figcaption{Results of the simulations with the accretion break, $r_{b1}=2$ km at 50 Myr but varying the density. Top: \textbf{Solid line} - nominal parameters with $45 \mbox{ M$_\oplus$}$, \textbf{Dashed line} - nominal parameters with $4.5 \mbox{ M$_\oplus$}$.  Bottom: ratio of the two simulations presented above.  \label{fig:fig3}}   
\end{figure}

Since the probability of a single object having a collision is approximately linearly dependent on the density of particles, the rate at which the divot forms is also linearly dependent on the density of particles (see Equations~\ref{eq:Ncoll} and \ref{eq:tdiv}). Minimal variation in the results occur due to order of magnitude variations in the belt density at equivalent points in the simulations. This is demonstrated in Figure~\ref{fig:fig3}, where we compare the nominal simulation with a simulation using the same parameters with the density dropped by a factor of $10$.

\begin{figure}[h] %  figure placement: here, top, bottom, or page
   \centering
   \plotone{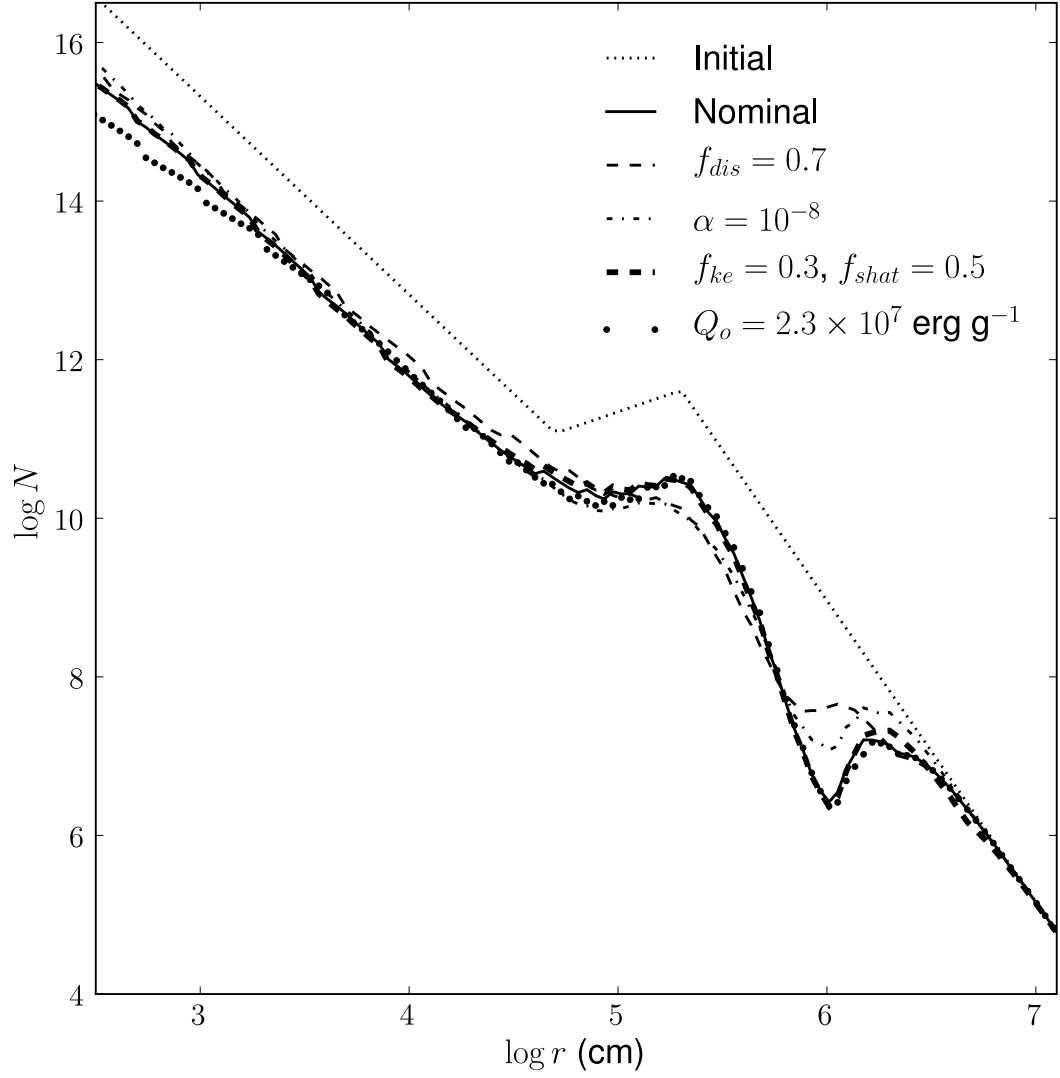} 
   \figcaption{Results of the simulations with the accretion break, $r_{b1}=2$ km at 500 Myr but varying parameters relevant to cratering. \label{fig:fig4}}   
\end{figure}

Presented in Figure~\ref{fig:fig4} are the results of our nominal simulations when the parameters relevant to the cratering model are changed. With certain strength models, namely those where cratering is responsible for removing a large fraction of object with $r\gtrsim 10 r_{b1}$,  small waves can form at those sizes. But for the majority of the simulations, the roll-over is smooth up to the edge of the divot where the density of objects decreases very rapidly. Other than the maximum cratering radius parameter $f_{dis}$ which affects the divot location, little variation in the size distribution for $r>r_{b2}$ results from variations in the cratering parameters. Changes do occur for $r<r_{b2}$. As there currently exists no measurement of the shape of the Kuiper belt size distribution in this size range, we leave the study of these parameters to later works. 

\section{Discussion \label{sec:discussion}}
The divot produced in our calculations could provide a natural explanation for the observed roll-over at $r\sim 25-60$ km detected by \citet{Bernstein2004} and confirmed by \citet{Fuentes2008} and \citet{Fraser2009}. That is, the size-distribution deviates away from the steep large-object accretion size-distribution into a divot at $r_{div}\sim 10-15$ km. If the observed roll-over is caused by a break left-over from accretion, then, from Equation~\ref{eq:divot}, we find that the accretion break must have occurred at a radius, $r_{b1}\sim0.5 - 3.5$ km, for our adopted strength scaling slope $b=1.19$. The implied accretion size break $r_{b1}$ is consistent with that found in previous accretion models \citep{Kenyon2002}.

Comparison of the ratio of objects larger and smaller than the kink radius, $r_k$, from the simulations, to the roll-over observed in the Kuiper belt can provide constraint on the mass of the belt at the end of accretion. Assuming that the accretion break-slope is, $q_2\sim0$, and that the Kuiper belt strength scaling is well represented by our assumed value $b=1.19$, a $45 \mbox{ M$_\oplus$}$ Kuiper belt would produce a divot consistent with observations in $\gtrsim 40$ Myr if $r_{b1}=2$ km and $\gtrsim 100$ Myr for $r_{b1}=1 \mbox{ km}$. $r_{b1}=2$. The divot formation timescales are a factor of $\sim2$ longer for the initial accretion break slope $q_2=2$ and $\sim 2$ shorter for the scaling law of \citet{Leinhardt2009}, ie. if objects are a factor of 3 weaker than suggested by \citet{Benz1999}. The time estimates also scale linearly with the density of planetesimals at the epoch of the dynamical excitation. 

Our Kuiper belt mass estimates are consistent with accretion simulations which suggest that the primordial Kuiper belt had a mass $\gtrsim 30 \mbox{ M$_{\oplus}$}$ \citep{Stern1997a,Kenyon1998,Kenyon2002}. If we assume that the mass depletion could not have occurred more than 1 Gyr after the Kuiper belt was excited, then the minimum Kuiper belt mass required to produce a divot compatible with the observations is $\sim1-5 \mbox{ M$_\oplus$}$.

The simulations of \citet{Charnoz2007} are the first attempt at coupling the dynamical and collisional history of the Kuiper belt. Using a similar initial mass distribution ($45 \mbox{ M$_\oplus$}$ between 5 and 50 AU) as we have adopted in our simulations, they found that the mass depletion which has produced the low-density modern day Kuiper belt must have been a result of combined collisional grinding and dynamical effects. Otherwise, the planetesimal populations external to the Kuiper belt - the Oort cloud and scattered disk \citep[see][]{Gladman2008} - would be too anemic by at least an order of magnitude. They found that the mass depletion of the Kuiper belt occurred over $\sim 100$ Myr. They however adopted an initial size distribution that broke from a steep large-object slope to that of collisional equilibrium rather than the accretion model inspired distribution we adopt here, complicating comparison of the two works. Roughly speaking however, the mass depletion timescale they found would be sufficient to produce a divot, if starting from an accretion-like size distribution. They also found that populations external to the Kuiper belt underwent significantly enhanced collisional evolution compared to the Kuiper belt. This suggests that the scattered disk and Oort cloud populations would have very few objects with $ 1 \lesssim r \lesssim 20$ km.

%A possible confirmation of this process might be detected in asteroid belt, which exhibits a kink in the size distribution at $r\sim 60$ km and a turn-up at $r\sim 6$ km. Assuming that this kink and turn-up pair are caused by a roll-over in the accretion size distribution for the asteroids, the roll-over should occur at $\sim0.2-0.5$ km (assuming a the basalt strength law of \citet{Benz1999} and the collision velocities found by \citet{Bottke2005}). Observations need to be made sensitive to such small asteroid sizes before the existence of the accretion roll-over can be confirmed.

Our results are consistent with some of the peculiarities of detected Kuiper belt objects. Only one object with radius $r\leq 15 \mbox{ km}$ has been detected \citep{Fuentes2009}. Such an observation is consistent with a flat $q\sim0$ power-law for objects smaller than the current roll-over. It is suggestive however, that there is a substantial depletion of objects with similar size. This is consistent with the idea that the size distribution is not a power-law for $r\sim10-20$ km, but rather has a shape like the divot produced in our calculations. The shape of the luminosity function presented by \citet{Fuentes2009} hints that a divot is detected at $m(R)\sim26.5$. The observational evidence of this however, is extremely sparse. More observations of small $r\sim10-40$ km Kuiper belt objects are required before the true size distribution shape is known.

We find that objects with sizes similar to the initial accretion break $r_{b1}$ do not suffer significant depletion on timescales necessary to form the divot. Our findings demonstrate that the size distribution should exhibit a depletion of objects for $r\sim 10$ km compared to objects with $r\sim r_{b1}$. Comparison of our results to that of \citet{Benavidez2009} demonstrates that the shape of the size distribution caused by collisional evolution is highly dependent on the initial size distribution; the break can be formed with different initial conditions, but the location, and interpretation of the break change. Our results demonstrate that the detection of a divot in the size distribution is distinct evidence for the existence of a break in the size distribution at the onset of collisional disruption.

Our calculations demonstrate that, with a detection of the remaining peak at $r_{b1}$, and the divot center $r_{div}$, the strength scaling slope $b$ as well as the accretion break radius $r_{b1}$ can be inferred. This would provide a stringent test of numerical impact simulations of icy-bodies, as well as a strong constraint on accretion models of the Kuiper belt region. The current uncertainty on the roll-over size cannot place anymore constraint on accretion or shattering physics of icy planetesimals, other than to say that the observations are consistent with the current understanding of these two processes.

Our findings suggest that the sky density of objects with $r\sim 1$ km is $\sim10^6-10^7$ objects per square degree. Detection by serendipitous occultation of star light is currently the only method of detection sensitive to these small KBOs. These exciting experiments have currently placed an upper limit of $\sim10^8-10^9$ objects per square degree at this size  \citep{Bickerton2008,Zhang2008} and future experiments will ultimately determine if the roll-over is caused by the mechanism we have suggested here. 

Our calculations, and the existence of the steep slope observed for the large object size distribution, suggests that the Kuiper belt underwent a short period of accretion, where objects as large as Eris were produced. Some dynamical event truncated accretion by increasing interaction velocities to disruption.  On a short $\sim 10-100$ Myr timescale, a divot, or depletion of objects with $r\sim 10$ km was produced before the dynamical scattering removed most of the mass from the system, freezing the shape of the size distribution, which has undergone minimal evolution to the current day.

\section{Conclusions \label{sec:conclusions}}
Our calculations have provided the first direct link from the ``typical'' size distribution produced from accretion, to the collisionally modified size distribution of the modern day Kuiper belt. Our calculations have demonstrated that, given a size distribution similar to that produced by accretion, the likely result of collisional evolution in the Kuiper belt, is to produce a divot in the Kuiper belt size distribution at $r\sim10-15$ km, and a roll-over from the large object accretion slope at $r\sim25-60$ km, compatible with the observed roll-over in the Kuiper belt size distribution. We conclude that the size distribution is likely not a power-law for objects smaller than the roll-over, and exhibits a dearth of objects at the divot location. Assuming a divot has been formed, then a measurement of the divot center radius would provide constraint on the strength scaling law in the gravity regime for KBOs, and constrain the size distribution of $r\sim 1$ km objects left-over from the accretion processes in the early Solar system.

From our calculations, we find that the Kuiper belt must have had at least $1-5 \mbox{ M$_\oplus$}$ of material after its objects were excited onto orbits with eccentricities comparable to those observed, otherwise the roll-over would not be produced on Gyr timescales. If the belt had a mass of $\sim 45 \mbox{ M$_\oplus$}$, then a divot would form in $\gtrsim 40$ Myr.

From the results of our calculations, we predict a deep absence of objects at $r\sim10$ km. We also predict  that the sky density of objects with $r\sim1$ km is $\sim10^6-10^7$ per square degree, 2-3 orders of magnitude larger than if the size distribution is a shallow power-law for objects smaller than the roll-over, as suggested by \citet{Fraser2009}.

\section{Acknowledgements}
A special thanks goes to JJ Kavelaars and Mike Brown, without who's guidance, this work would likely have never come to fruition. I would also like to thank David Trilling for his useful advice on what I should be looking for. Finally I would like to thank Parker, Raggozzine, Schwamb, and Bannister for humoring me during my rants.

This project was funded in part, by the National Science and Engineering Research Council and the National Research Council of Canada. Support for this work was provided in part by NASA through a grant from the program HST-GO-11644 from the Space Telescope Science Institute, which is operated by the Association of Universities for Research in Astronomy, Inc. under NASA contract NAS 5-26555.  This research used the facilities of the Canadian Astronomy Data Centre operated by the National Research Council of Canada with the support of the Canadian Space Agency.

\appendix
\section{Appendix A \label{sec:AppendixA}}
Here we present various tests to confirm the correct behavior of the collisional model. As done by \citet{Ohtsuki1990} and \citet{Wetherill1990}, the behavior of a collisional evolution model -at least in terms of accretion - can be assessed by comparing numerical simulations of coagulation to the analytic solutions of the discrete coagulation equation given by

\begin{equation}
\frac{d}{dt} n_k = \frac{1}{2} \sum_{i+j=k} A_{ij}n_i n_j - n_k \sum_{i=1} A_{ik} n_i.
\label{eq:coag}
\end{equation}

\noindent Here, $\frac{d}{dt} n_k$ is the time rate of change of objects in bin $k$ from collisions adding objects to (left term) and removing objects from (right term) bin $k$ summed over all bins. $A_{ij}$ is the ``kernel'', and governs the probability of collisions between $i$ and $j$. This is usually written in terms of mass bins $i$ and $j$ rather than the size bins we consider above. Any successful accretion or collision model must reproduce analytic solutions of Equation~\ref{eq:coag}. Discrepancies between the numerical and analytic solutions can highlight possible caveats about the results of the numerical models. \citet{Ohtsuki1990} has demonstrated that, in his constant mass bin formalism, the numerical solutions are typically accelerated compared to the analytic counter parts. It is worth noting that, in the mass-batch formalism of \citet{Wetherill1990}, the numerical solution {\it lags} the analytic one. As we use the constant mass bin formalism, the conclusions about the behaviour of his model are applicable here. We discuss this below. The acceleration decreases with the bin-spacing $\delta$,\footnote{As done in the main text, we still refer to $\delta$ here in terms of object size. Our $\delta$  is the cube of that used in the coagulation equation solution.} and varies with each kernel used. Thus, comparison of the numerical and analytic solution can provide a metric for the largest allowable $\delta$. These comparisons also provide a measure of how large a time-step can be used before the incorrect evolution is produced.

The simplest kernel is a constant, ie. $A_{ij}=\lambda=constant$, ie. the collision probability doesn't depend on target mass. This kernel produces smooth orderly growth, and the solution to the coagulation equation for this kernel, ie. the number of objects in a bin, $n_k$,  is $\frac{n_k}{n_o}=f^2 \left(1-f\right)^{k-1}$ where $f=\left(1+\eta/2\right)^{-1}$ and $\eta=\lambda n_o t$ is the dimensionless time. $n_o$ is the total number of objects at time zero, all of which are of the smallest mass. 

\begin{figure}[h] %  figure placement: here, top, bottom, or page
   \centering
   \plotone{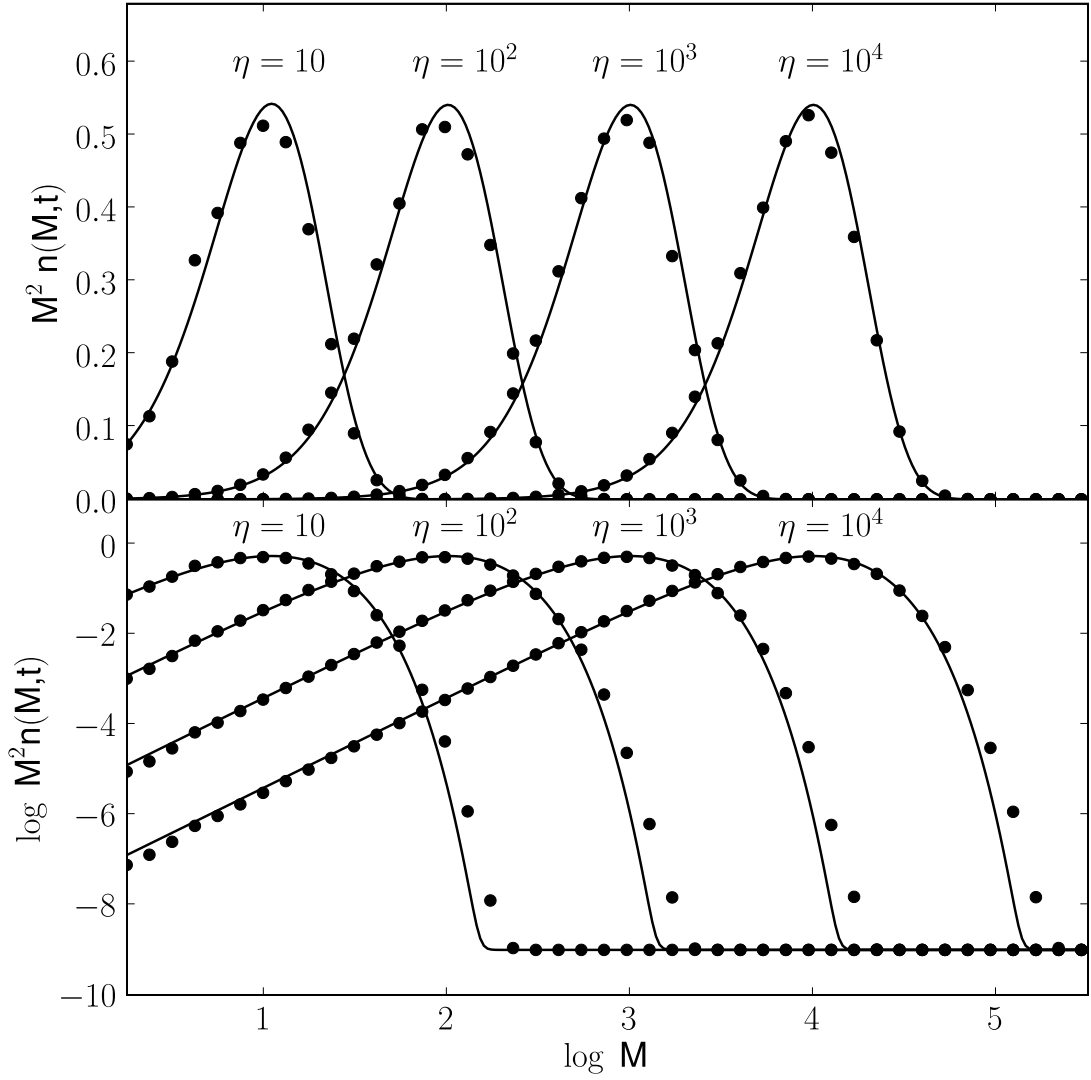} 
   \figcaption{ Analytic (solid line) and numerical (points) solutions to the coagulation equation at different time intervals for the kernel $A=\Delta$. \label{fig:figA1}}   
\end{figure}

The analytic solution and the numerical solution of the coagulation equation with the constant kernel are presented in Figure~\ref{fig:figA1} utilizing a $\delta=1.1$, $n_o=10^{18}$ particles\footnote{it was found that for $n_o\gtrsim10^{10}$, the results were independent of the choice in $n_o$}, 128 bins, and by not allowing the number of objects in a bin to change by more than $0.1 \%$ per time-step. As can be seen, the numerical solution matches the analytic solution well. The discrepancies present are similar to those found by \citet{Ohtsuki1990}. The numerical solution produces more large objects than in the analytic one. The difference however, is almost negligible and increases for larger $\delta$.

Another kernel for which an analytic solution exists is for collision probabilities proportional to the sum of colliding masses, ie. $A_{ij}=\beta\left(i+j\right)$. The analytic solution for this kernel is $n_k=n_o \frac{k^{k-1}}{k!} f\left(1-f\right)^{k-1}e^{-k(1-f)}$ where $f=e^{-\eta}$ and $\eta=\beta n_o t$ is the dimensionless time. The analytic and numerical solutions for this kernel are presented in Figures~\ref{fig:figA2} and \ref{fig:figA3} for $n_o=10^{18}$ particles, 128 bins, and $\delta=1.08.$\footnote{chosen to ensure the largest mass bin is larger than the largest object in the analytic solution.} This kernel creates large objects more rapidly than the constant kernel. As such, the numerical solution is more sensitive to the choice in $\delta$ than for the constant kernel. As shown by \citet{Ohtsuki1990}, a finite $\delta$ in the numerical solution creates an acceleration in the coagulation, as well as an over-abundance of large objects compared to the analytic solution. Our model exhibits virtually identical behavior. We find that the acceleration is nearly a constant over the simulations (see Figure~\ref{fig:figA3}). That is, for $\delta=1.08$ the shape of the numerical solution best matches the analytic solution at a time $\sim12\%$ earlier than expected. This factor is constant over the simulations, and scales with the choice in $\delta$. 

The kernel proportional to the product of the colliding masses, ie. $A_{ij}=\gamma \left(i*j\right)$ produces runaway growth at $\eta=1$ (again $\eta=\gamma n_o t$ is the dimensionless time). As such, the kernel is difficult to solve analytically, as the time-step must vary with mass of the run-away body to re-create the proper evolution. The analytic solution for the non-runaway bodies is $n_k=\frac{n_o \left(2k\right)^{k-1}}{k!k}\left(\frac{\eta}{2}\right)^{k-1}e^{-k\eta}$, and the mass of the runaway body is $m=n_o \exp \left[-\int \sum_{k=1}^{N} k^2 \frac{n_k}{n_o} d\eta'\right]$ \citep{Wetherill1990}.

The code presented here was intended for collisional evolution calculations, where the primary process is disruption of objects to ever smaller sizes. The code inefficiently handles the case of runaway growth; the code takes many more steps than necessary, causing the runtime to increase substantially as runaway continues. Our code was run only long enough to demonstrate that the numerical solution produces the correct mass of the runaway body, and distribution of smaller bodies (see Figure~\ref{fig:figA4}). Future versions of the software will include modifications to handle runaway growth more efficiently. But as the simulations presented above have minimal accretion of even the largest objects, inclusion of these modifications was unnecessary, and omitted for this work.

Presented in Figure~\ref{fig:figA4} is a comparison of the numerical and analytic solutions to the coagulation equation before and after runaway begins using $n_o=10^{18}$ and $\delta=1.06$, and allowing the number of objects per bin to vary by 0.1\% per time-step. Much like the previous examples, the numerical solution produces an accelerated evolution compared to the analytic solution. The acceleration in the numerical solution of runaway growth is not constant with time, as is found in other numerical models \citep[see for instance][]{Kenyon1998}. The acceleration is most pronounced for short times $(\sim 20\% \mbox{ when } \eta=0.4)$, and decreases with time. 

In the numerical solution, runaway growth occurs at $\eta=0.976$, slightly early compared to the analytic solution. The post-runaway numerical solutions are presented in Figure~\ref{fig:figA4} at $\eta=0.98$ and 1.01, and very accurately match the analytic solutions at $\eta=1.0003$ and 1.0006 respectively. The mass of the runaway body in the numerical solution at these two times, is $3.5\times10^{-4}$ and $5.6\times10^{-4}$ the total system mass. The mass of the runaway body in the analytic solution at the times where the analytic and numerical solutions match is $3.5\times10^{-4}$ and $6\times10^{-4}$ the total system mass, almost identical to the numerical solution. The results presented in Figure~\ref{fig:figA4} demonstrate that our code accurately reproduces the analytic solution to the coagulation equation for the case of runaway growth, before and after runaway commences.

The solution to this kernel is much more sensitive to the choice in $\delta$. For $\delta\geq 1.1$, the acceleration verges on unacceptably large, and the mass of the run-away body is not correctly reproduced. Therefore, $\delta=1.1$ is an approximate upper-limit on the bin spacing for simulations in which runaway growth occurs, and which utilize our model.

\section{Appendix B \label{sec:AppendixB}}
Here we present tests of the collisional evolution considering various details of our model. Specifically we test the effects of time-step on our collisional evolution calculations and the numerical solutions to the coagulation equation. Additionally, we test our adopted collisional equilibrium forcing over the smallest bins.

\begin{figure}[h] %  figure placement: here, top, bottom, or page
   \centering
   \plotone{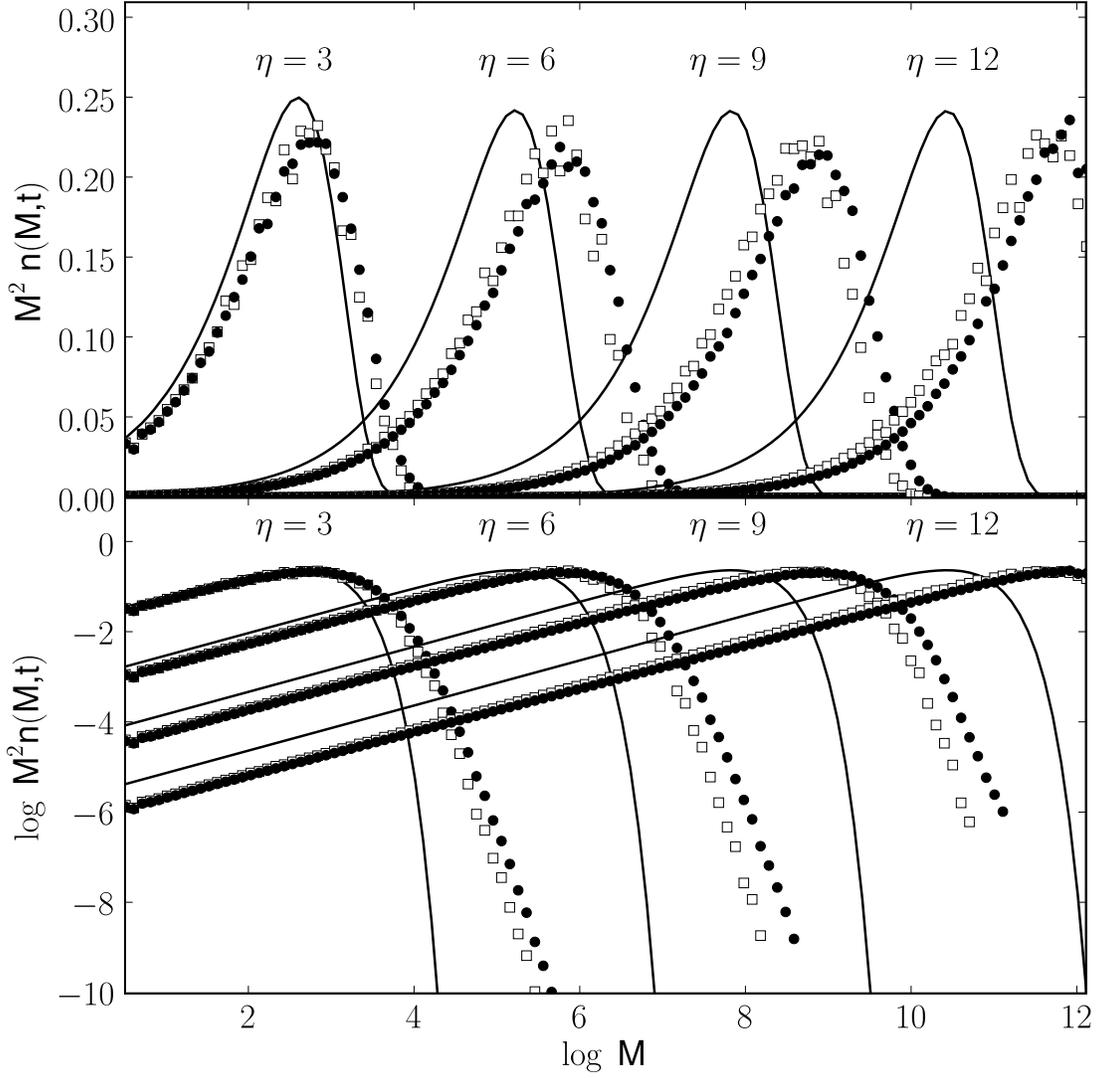} 
   \figcaption{ Analytic (solid line) and numerical (points) solutions to the coagulation equation at different time intervals for the kernel $A=\beta(i+j)$. Dots and squares represent results when the bins are allowed to vary by 0.1\% and 1\%  respectively. \label{fig:figA2}}   
\end{figure}

\begin{figure}[h] %  figure placement: here, top, bottom, or page
   \centering
   \plotone{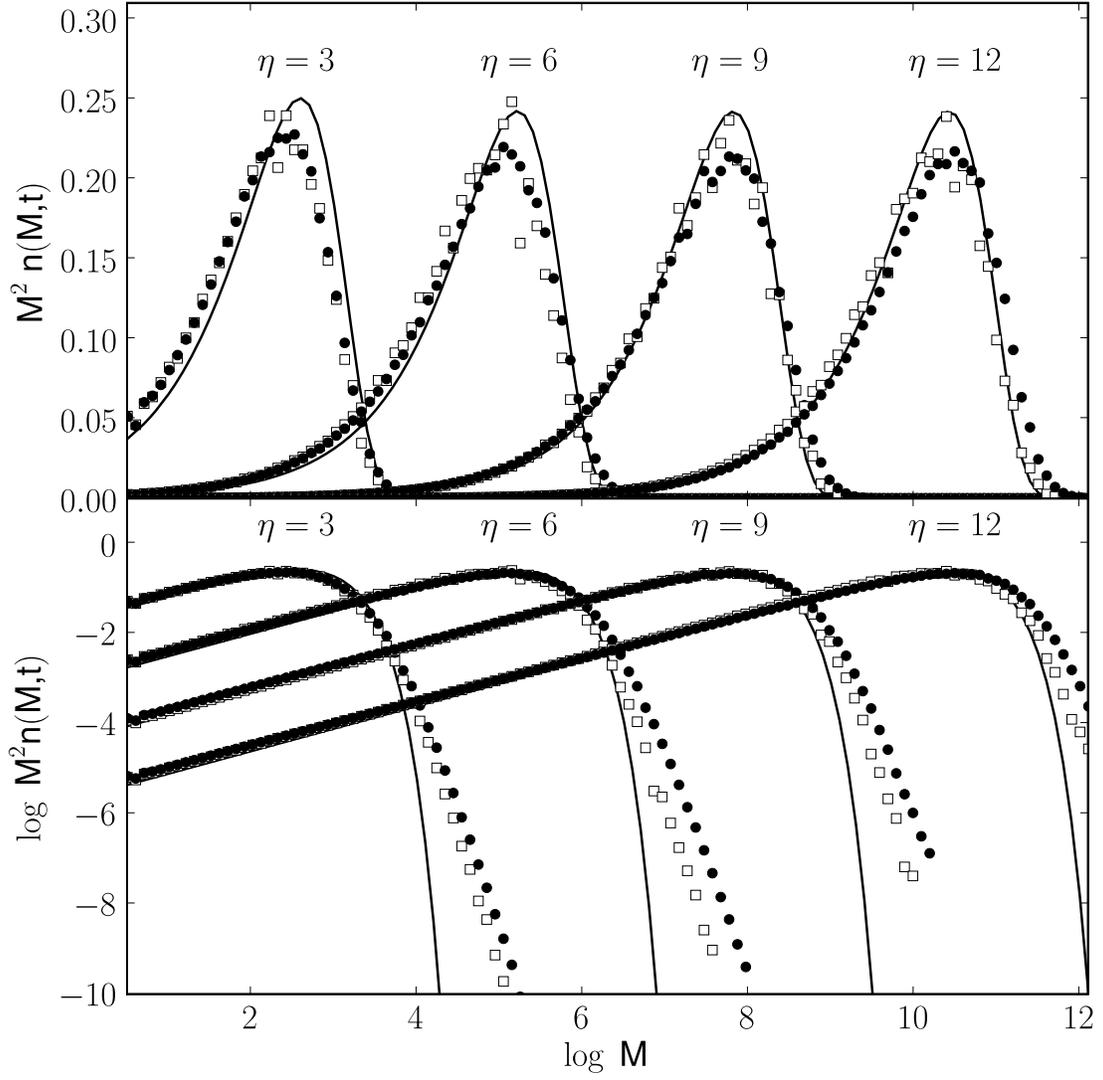} 
   \figcaption{ As in Figure~\ref{fig:figA2}, but with the numerical solution presented at a time 12\% earlier than the analytic solution. \label{fig:figA3}}   
\end{figure}

\begin{figure}[h] %  figure placement: here, top, bottom, or page
   \centering
   \plotone{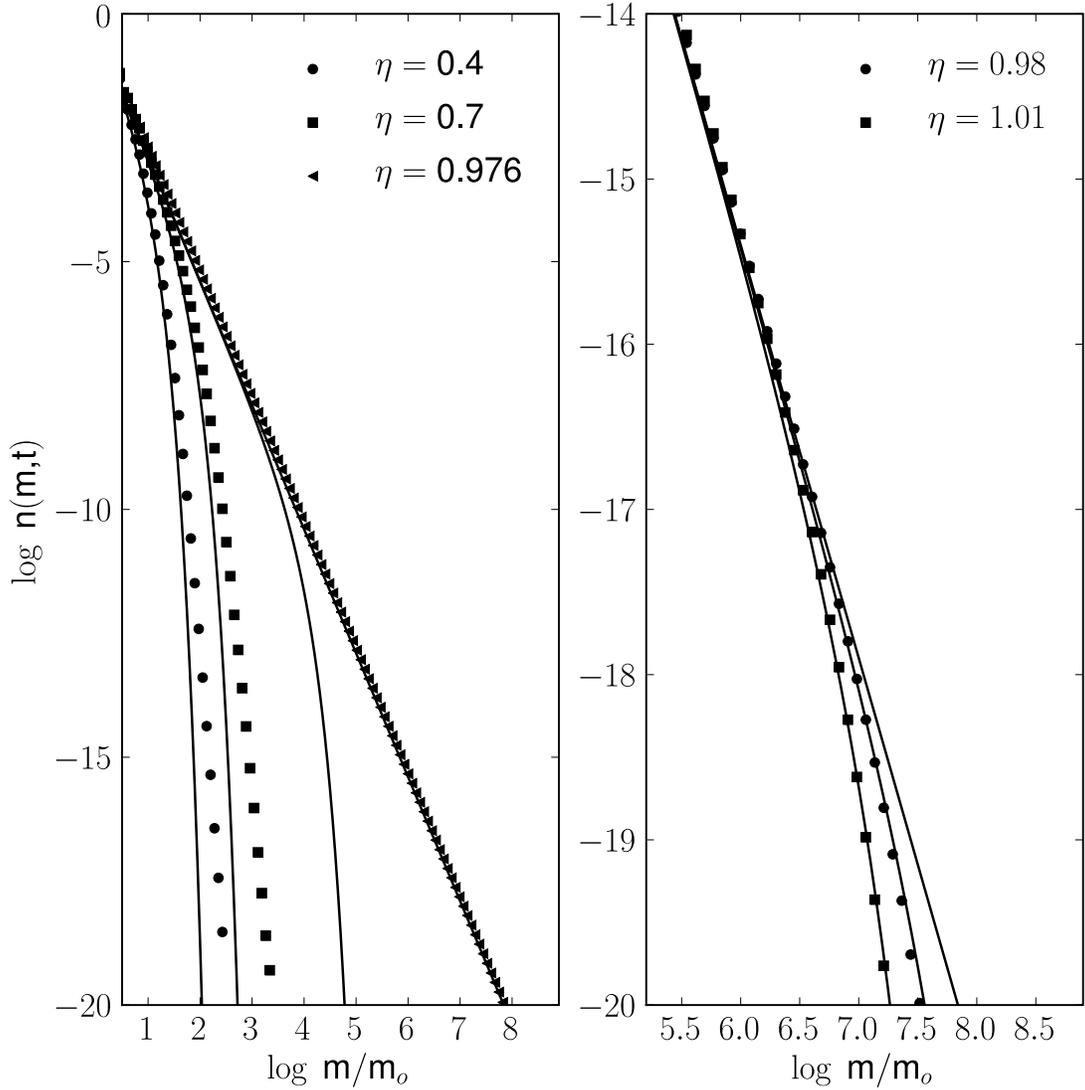} 
   \figcaption{ Analytic (solid lines) and numerical (points) solutions to the coagulation equation at different time intervals for the kernel $A=\gamma(ij)$. Left: Before runaway growth commences. Presented are numerical solutions at $\eta=0.4$, 0.7, and 0.976, along with analytic solutions at $\eta=0.4$, 0.7, 0.976, and 1. Right: Snap shot of the large non-runaway bodies, where the evolution is most rapid, after runaway growth commences. Presented are numerical solutions at $\eta=0.98$ and 1.01 along with analytic solutions at $\eta=1.0003$ and 1.0006. \label{fig:figA4}}   
\end{figure}

\begin{figure}[h] %  figure placement: here, top, bottom, or page
   \centering
   \plotone{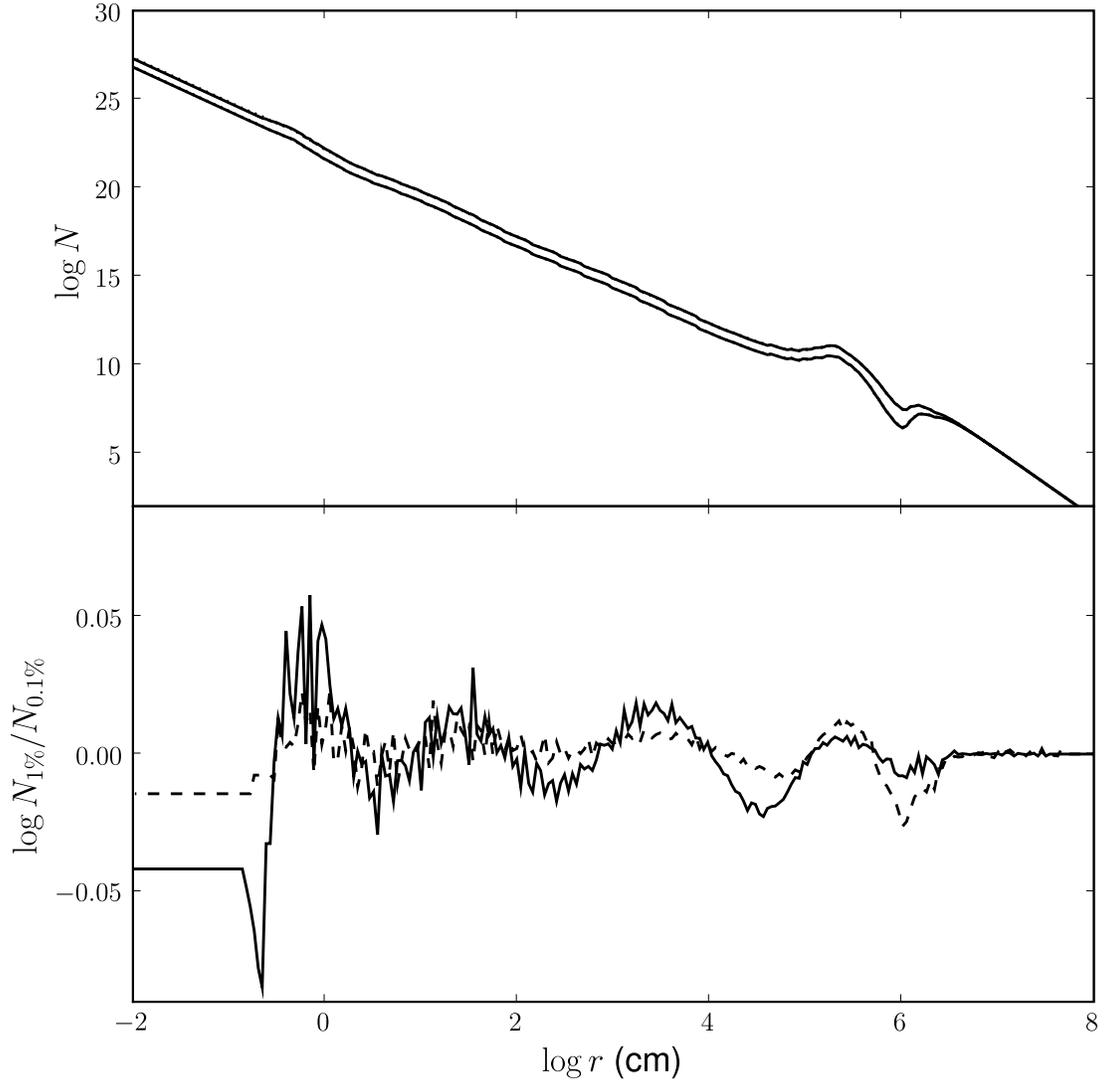} 
   \figcaption{Top: Results of collisional evolution for the nominal parameters at 100 and 500 Myr when allowing the number of objects in a bin to change by 0.1\% (solid) and 1\% (dotted) per timestep. Bottom: Ratio of the simulations presented in the top using two separate timesteps at 100 (dashed) and 500 (solid) Myr. \label{fig:figA5}}   
\end{figure}

The choice in time-step modifies the results of the calculations; too large and the results will not accurately produce the correct evolution. While large time-steps may repeat the general behavior of simulations which use smaller time-steps, significant stochastic variations in the results occur if the time-step used is too large. It was found  that not allowing the number of objects in a bin to change by more than $1 \%$ per time-step was sufficient to ensure accuracy in our collisional evolution simulations as well as the solutions to the coagulation equation for kernels $A=\Delta$ and $A=\beta(i+j)$ (see Figures~\ref{fig:figA2}, \ref{fig:figA3}, and \ref{fig:figA5}). Further tightening this condition did not change the results by more than a few percent, while relaxing this condition further produced incorrect results for all numerical solutions. Note: none of the numerical solutions of the coagulation equation include a probabilistic treatment of the collision rates as given in Equation~\ref{eq:Ndis}. Therefore, these solutions suffer the most from larger time-steps than do the main simulations presented here.

\begin{figure}[h] %  figure placement: here, top, bottom, or page
   \centering
   \plotone{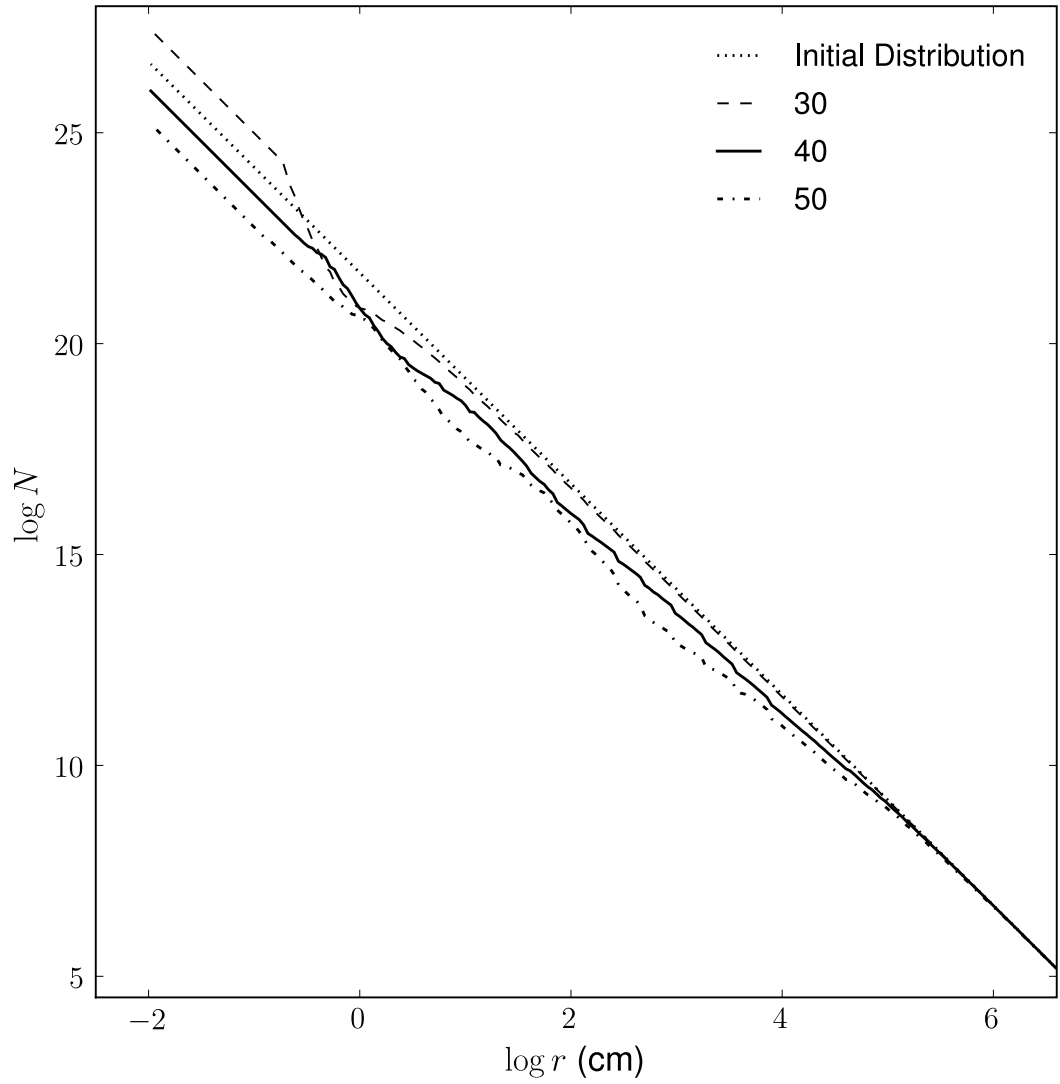} 
   \figcaption{Results of collisional evolution after 2000 time-steps, starting from a distribution in collisional equilibrium (dotted line). The various lines are the results when equilibrium was forced over 30, 40, and 50 bins.  \label{fig:figA6}}   
\end{figure}

For the simulations in which collisional equilibrium was forced for the smallest bins, we tested over how many bins this should be applied. The results are shown in Figure~\ref{fig:figA6} where we present the evolution of a system already in collisional equilibrium, but forcing equilibrium on the smallest 30, 40, and 50 bins. 

If equilibrium is forced over 30 bins only, a large mass build-up occurs at the small-size end. This build-up increases in magnitude over time, and produces unrealistic results. If collisional equilibrium was forced over 40 and 50 bins, the mass build-up no longer occurs, and the smallest object become collisionally depleted, as expected. The minimum number of bins required to prevent the mass build up varies with the cratering and fragment distributions chosen for the simulations. The rate of depletion does depend on the number of bins over which collisional equilibrium is forced. The results however, are similar for objects more than $\sim2$ orders of magnitude larger than the smallest bin. This effect will not affect the main conclusions drawn above about the large object size distribution. 

As demonstrated by \citet{Thebault2003}, proper treatment of the smallest objects requires modeling of the radiation forces which are ultimately responsible for removing the dust out of the region. Such a treatment is beyond the scope of this work. We choose to force collisional equilibrium over the smallest 40 bins, as a compromise between having an unrealistic small object mass buildup, and potentially removing mass too quickly from the system. All conclusions drawn above have the caveat, that the small object size distribution might not be correctly accounted for here.

%\bibliographystyle{apj}
%\bibliography{AstroElsart}

\begin{thebibliography}{41}
\expandafter\ifx\csname natexlab\endcsname\relax\def\natexlab#1{#1}\fi

\bibitem[{{Benavidez} \& {Campo Bagatin}(2009)}]{Benavidez2009}
{Benavidez}, P.~G. \& {Campo Bagatin}, A. 2009, \planss, 57, 201

\bibitem[{{Benz} \& {Asphaug}(1999)}]{Benz1999}
{Benz}, W. \& {Asphaug}, E. 1999, Icarus, 142, 5

\bibitem[{{Bernstein} {et~al.}(2004){Bernstein}, {Trilling}, {Allen}, {Brown},
  {Holman}, \& {Malhotra}}]{Bernstein2004}
{Bernstein}, G.~M., {Trilling}, D.~E., {Allen}, R.~L., {Brown}, M.~E.,
  {Holman}, M., \& {Malhotra}, R. 2004, AJ, 128, 1364

\bibitem[{{Bickerton} {et~al.}(2008){Bickerton}, {Kavelaars}, \&
  {Welch}}]{Bickerton2008}
{Bickerton}, S.~J., {Kavelaars}, J.~J., \& {Welch}, D.~L. 2008, \aj, 135, 1039

\bibitem[{{Bottke} {et~al.}(2005){Bottke}, {Durda}, {Nesvorn{\'y}}, {Jedicke},
  {Morbidelli}, {Vokrouhlick{\'y}}, \& {Levison}}]{Bottke2005}
{Bottke}, W.~F., {Durda}, D.~D., {Nesvorn{\'y}}, D., {Jedicke}, R.,
  {Morbidelli}, A., {Vokrouhlick{\'y}}, D., \& {Levison}, H. 2005, Icarus, 175,
  111

\bibitem[{{Brown}(2001)}]{Brown2001}
{Brown}, M.~E. 2001, AJ, 121, 2804

\bibitem[{{Campo Bagatin} {et~al.}(1994){Campo Bagatin}, {Cellino}, {Davis},
  {Farinella}, \& {Paolicchi}}]{CampoBagatin1994}
{Campo Bagatin}, A., {Cellino}, A., {Davis}, D.~R., {Farinella}, P., \&
  {Paolicchi}, P. 1994, \planss, 42, 1079

\bibitem[{{Charnoz} \& {Morbidelli}(2007)}]{Charnoz2007}
{Charnoz}, S. \& {Morbidelli}, A. 2007, Icarus, 188, 468

\bibitem[{{Davis} \& {Farinella}(1997)}]{Davis1997}
{Davis}, D.~R. \& {Farinella}, P. 1997, Icarus, 125, 50

\bibitem[{{Dell'Oro} {et~al.}(2001){Dell'Oro}, {Marzari}, {Paolicchi}, \&
  {Vanzani}}]{Delloro2001}
{Dell'Oro}, A., {Marzari}, F., {Paolicchi}, P., \& {Vanzani}, V. 2001, \aap,
  366, 1053

\bibitem[{{Dohnanyi}(1969)}]{Dohnanyi1969}
{Dohnanyi}, J.~W. 1969, J.~Geophys.~Res., 74, 2531

\bibitem[{{Durda} \& {Stern}(2000)}]{Durda2000}
{Durda}, D.~D. \& {Stern}, S.~A. 2000, Icarus, 145, 220

\bibitem[{{Fraser} \& {Kavelaars}(2009)}]{Fraser2009}
{Fraser}, W.~C. \& {Kavelaars}, J.~J. 2009, \aj, 137, 72

\bibitem[{{Fraser} {et~al.}(2008){Fraser}, {Kavelaars}, {Holman}, {Pritchet},
  {Gladman}, {Grav}, {Jones}, {Macwilliams}, \& {Petit}}]{Fraser2008}
{Fraser}, W.~C., {Kavelaars}, J.~J., {Holman}, M.~J., {Pritchet}, C.~J.,
  {Gladman}, B.~J., {Grav}, T., {Jones}, R.~L., {Macwilliams}, J., \& {Petit},
  J.-M. 2008, Icarus, 195, 827

\bibitem[{{Fuentes} {et~al.}(2009){Fuentes}, {George}, \&
  {Holman}}]{Fuentes2009}
{Fuentes}, C.~I., {George}, M.~R., \& {Holman}, M.~J. 2009, \apj, 696, 91

\bibitem[{{Fuentes} \& {Holman}(2008)}]{Fuentes2008}
{Fuentes}, C.~I. \& {Holman}, M.~J. 2008, \aj, 136, 83

\bibitem[{{Gladman} {et~al.}(2001){Gladman}, {Kavelaars}, {Petit},
  {Morbidelli}, {Holman}, \& {Loredo}}]{Gladman2001}
{Gladman}, B., {Kavelaars}, J.~J., {Petit}, J.-M., {Morbidelli}, A., {Holman},
  M.~J., \& {Loredo}, T. 2001, AJ, 122, 1051

\bibitem[{{Gladman} {et~al.}(2008){Gladman}, {Marsden}, \&
  {Vanlaerhoven}}]{Gladman2008}
{Gladman}, B., {Marsden}, B.~G., \& {Vanlaerhoven}, C. 2008, {Nomenclature in
  the Outer Solar System} (The Solar System Beyond Neptune), 43--57

\bibitem[{{Gomes}(2003)}]{Gomes2003}
{Gomes}, R.~S. 2003, Icarus, 161, 404

\bibitem[{{Kenyon}(2002)}]{Kenyon2002}
{Kenyon}, S.~J. 2002, PASP, 114, 265

\bibitem[{{Kenyon} \& {Bromley}(2001)}]{Kenyon2001}
{Kenyon}, S.~J. \& {Bromley}, B.~C. 2001, AJ, 121, 538

\bibitem[{{Kenyon} \& {Bromley}(2004)}]{Kenyon2004}
---. 2004, AJ, 128, 1916

\bibitem[{{Kenyon} {et~al.}(2008){Kenyon}, {Bromley}, {O'Brien}, \&
  {Davis}}]{Kenyon2008}
{Kenyon}, S.~J., {Bromley}, B.~C., {O'Brien}, D.~P., \& {Davis}, D.~R. 2008,
  {Formation and Collisional Evolution of Kuiper Belt Objects}, ed. M.~A.
  {Barucci}, H.~{Boehnhardt}, D.~P. {Cruikshank}, \& A.~{Morbidelli}, 293--313

\bibitem[{{Kenyon} \& {Luu}(1998)}]{Kenyon1998}
{Kenyon}, S.~J. \& {Luu}, J.~X. 1998, \aj, 115, 2136

\bibitem[{{Leinhardt} \& {Stewart}(2009)}]{Leinhardt2009}
{Leinhardt}, Z.~M. \& {Stewart}, S.~T. 2009, Icarus, 199, 542

\bibitem[{{Leinhardt} {et~al.}(2008){Leinhardt}, {Stewart}, \&
  {Schultz}}]{Leinhardt2008}
{Leinhardt}, Z.~M., {Stewart}, S.~T., \& {Schultz}, P.~H. 2008, {Physical
  Effects of Collisions in the Kuiper Belt}, ed. M.~A. {Barucci},
  H.~{Boehnhardt}, D.~P. {Cruikshank}, \& A.~{Morbidelli}, 195--211

\bibitem[{{Morbidelli} {et~al.}(2008){Morbidelli}, {Levison}, \&
  {Gomes}}]{Morbidelli2008}
{Morbidelli}, A., {Levison}, H.~F., \& {Gomes}, R. 2008, {The Dynamical
  Structure of the Kuiper Belt and Its Primordial Origin} (The Solar System
  Beyond Neptune), 275--292

\bibitem[{{Ohtsuki} {et~al.}(1990){Ohtsuki}, {Nakagawa}, \&
  {Nakazawa}}]{Ohtsuki1990}
{Ohtsuki}, K., {Nakagawa}, Y., \& {Nakazawa}, K. 1990, Icarus, 83, 205

\bibitem[{{Pan} \& {Sari}(2005)}]{Pan2005}
{Pan}, M. \& {Sari}, R. 2005, Icarus, 173, 342

\bibitem[{{Petit} \& {Farinella}(1993)}]{Petit1993}
{Petit}, J.-M. \& {Farinella}, P. 1993, Celestial Mechanics and Dynamical
  Astronomy, 57, 1

\bibitem[{{Petit} {et~al.}(2006){Petit}, {Holman}, {Gladman}, {Kavelaars},
  {Scholl}, \& {Loredo}}]{Petit2006}
{Petit}, J.-M., {Holman}, M.~J., {Gladman}, B.~J., {Kavelaars}, J.~J.,
  {Scholl}, H., \& {Loredo}, T.~J. 2006, MNRAS, 365, 429

\bibitem[{{Petit} {et~al.}(2008){Petit}, {Kavelaars}, {Gladman}, \&
  {Loredo}}]{Petit2008}
{Petit}, J.-M., {Kavelaars}, J.~J., {Gladman}, B., \& {Loredo}, T. 2008, {Size
  Distribution of Multikilometer Transneptunian Objects}, ed. M.~A. {Barucci},
  H.~{Boehnhardt}, D.~P. {Cruikshank}, \& A.~{Morbidelli}, 71--87

\bibitem[{{Stern}(1996)}]{Stern1996a}
{Stern}, S.~A. 1996, \aj, 112, 1203

\bibitem[{{Stern} \& {Colwell}(1997)}]{Stern1997a}
{Stern}, S.~A. \& {Colwell}, J.~E. 1997, \aj, 114, 841

\bibitem[{{Stewart} \& {Leinhardt}(2009)}]{Stewart2009}
{Stewart}, S.~T. \& {Leinhardt}, Z.~M. 2009, \apjl, 691, L133

\bibitem[{{Th{\'e}bault} {et~al.}(2003){Th{\'e}bault}, {Augereau}, \&
  {Beust}}]{Thebault2003}
{Th{\'e}bault}, P., {Augereau}, J.~C., \& {Beust}, H. 2003, \aap, 408, 775

\bibitem[{{Trujillo} {et~al.}(2001){Trujillo}, {Jewitt}, \&
  {Luu}}]{Trujillo2001b}
{Trujillo}, C.~A., {Jewitt}, D.~C., \& {Luu}, J.~X. 2001, AJ, 122, 457

\bibitem[{{Wetherill}(1990)}]{Wetherill1990}
{Wetherill}, G.~W. 1990, Icarus, 88, 336

\bibitem[{{Wetherill} \& {Stewart}(1989)}]{Wetherill1989}
{Wetherill}, G.~W. \& {Stewart}, G.~R. 1989, Icarus, 77, 330

\bibitem[{{Wetherill} \& {Stewart}(1993)}]{Wetherill1993}
---. 1993, Icarus, 106, 190

\bibitem[{{Zhang} {et~al.}(2008){Zhang}, {Bianco}, {Lehner}, {Coehlo}, {Wang},
  {Mondal}, {Alcock}, {Axelrod}, {Byun}, {Chen}, {Cook}, {Dave}, {de Pater},
  {Porrata}, {Kim}, {King}, {Lee}, {Lin}, {Lissauer}, {Marshall}, {Protopapas},
  {Rice}, {Schwamb}, {Wang}, \& {Wen}}]{Zhang2008}
{Zhang}, Z.-W., {Bianco}, F.~B., {Lehner}, M.~J., {Coehlo}, N.~K., {Wang},
  J.-H., {Mondal}, S., {Alcock}, C., {Axelrod}, T., {Byun}, Y.-I., {Chen},
  W.~P., {Cook}, K.~H., {Dave}, R., {de Pater}, I., {Porrata}, R., {Kim},
  D.-W., {King}, S.-K., {Lee}, T., {Lin}, H.-C., {Lissauer}, J.~J., {Marshall},
  S.~L., {Protopapas}, P., {Rice}, J.~A., {Schwamb}, M.~E., {Wang}, S.-Y., \&
  {Wen}, C.-Y. 2008, \apjl, 685, L157

\end{thebibliography}

%examples in case you need them
\begin{deluxetable}{cc}
   \tablecaption{Nominal Simulation Parameters \label{tab:params}}
   \startdata	
	Parameter & Adopted Value \\ \hline
	\multicolumn{2}{c}{Disruption Parameters}\\ \hline
	$Q_o$ & $7\times10^7\mbox{ erg g$^{-1}$}$\\
	$B$ & $2.1 \mbox { erg cm$^3$ g$^{-2}$}$ \\
	$a$ & $-0.45$ \\
	$b$ & $1.19$\\ 
	$X$ & $0.55$ \\ 
	$Y$ & $1$\\ \hline
	\multicolumn{2}{c}{Cratering Parameters}\\ \hline
	$\alpha$ & $10^{-9} \mbox{ s$^2$ cm$^{-2}$}$ \\
	$f_{KE}$ & $0.1$ \\
	$f_{frac}$ & $0.8$ \\
	$f_{Crat}$ & 0.05 \\
	$q_c$ & $3.4$ \\
	$k$ & $\frac{9}{4}$ \\ \hline
	\multicolumn{2}{c}{Miscellaneous Parameters}\\ \hline
	$\rho$ & $1.2 \mbox{ g cm$^{-3}$}$ \\
	$v_{rel}$ & $1 \mbox{ km s$^{-1}$}$\\ 
	$r_{min}$ & 0.01 cm \\
	$\delta$ & $1.1$ \\ \hline
	\multicolumn{2}{c}{Initial Size Distribution Parameters}\\ \hline
	$q_1$ & $4.8$ \\
	$q_2$ & $0-2$ \\
	$q_3$ & $3.5$ \\
	$r_{b1}$ & $2$ km\\
	$r_{b2}$ & $0.5$ km\\ 
		
   \enddata
\end{deluxetable}
   
\onecolumn

\end{document}